\documentclass[journal]{IEEEtran}

\usepackage{cite}
\usepackage{amsmath,amssymb,amsfonts}
\usepackage{algorithmic}
\usepackage{graphicx}
\usepackage{textcomp}
\usepackage{xcolor}
\usepackage{booktabs}
\usepackage{array}
\usepackage{multirow}
\usepackage{url}

\newcommand{\figplaceholder}[2]{%
  \fbox{\begin{minipage}[c][#1][c]{#2}\centering
  \vspace{0.5em}
  Placeholder for experimental figure
  \vspace{0.5em}
  \end{minipage}}%
}

\newcommand{\blueadd}[1]{\textcolor{black}{#1}}

\begin{document}

\title{Towards Industrial Application: Differentiable Predictive Current Control for PMSM Drives}

\author{Yuan~Li,~\IEEEmembership{Member,~IEEE,}
        Mateja~Novak,~\IEEEmembership{Member,~IEEE,}
        Shuai~Zhao,~\IEEEmembership{Senior Member,~IEEE,}
        Radu~Dan~Lazar,
        Norbert~Hanigovszki,
        Huai~Wang,~\IEEEmembership{Senior Member,~IEEE,}
        and~Frede~Blaabjerg,~\IEEEmembership{Fellow,~IEEE}%
\thanks{This work was supported by Innovation Fund Denmark through the Artificial Intelligence for Next-Generation Power Electronics project.}
\thanks{Yuan Li, Mateja Novak, Shuai Zhao, Huai Wang, and Frede Blaabjerg are with AAU Energy, Aalborg University, Aalborg, Denmark (e-mail: yuanli@energy.aau.dk).}
\thanks{Radu Dan Lazar and Norbert Hanigovszki are with Danfoss Drives A/S, Gråsten, Denmark.}}

\markboth{IEEE Transactions on Power Electronics,~Vol.~XX, No.~XX, Month~2026}%
{Li \MakeLowercase{\textit{et al.}}: Differentiable Predictive Current Control for PMSM Drives}

\maketitle

\begin{abstract}
Model predictive control (MPC) is attractive for high-performance permanent magnet synchronous motor (PMSM) drives because of its fast dynamic response and systematic handling of multivariable control objectives. However, solving a predictive optimization problem at every sampling instant can impose a considerable computational burden on industrial embedded controllers with short control periods. To address this issue, this paper proposes a differentiable predictive current control (DPCC) strategy as a lightweight neural-network-based alternative to online-optimized MPC. Instead of learning from precomputed optimal control labels, the proposed DPCC embeds the PMSM prediction model and the predictive current-tracking objective into a differentiable computational graph, allowing the control policy to be trained directly offline in a label-free manner. During real-time operation, the trained policy generates the voltage command through a single forward inference, thereby eliminating iterative online optimization. The proposed DPCC is implemented and experimentally evaluated on an industrial PMSM drive platform at Danfoss Drives within a practical sensorless drive-control architecture. Tests under current transients, load-torque disturbances, inductance mismatch, and steady-state conditions demonstrate that DPCC achieves performance comparable to an optimized MPC benchmark. Meanwhile, DPCC reduces the current-controller execution time to 3.5~$\mu$s, representing a 79\% reduction compared with MPC. These results demonstrate the practical feasibility of DPCC for computationally efficient real-time implementation in industrial PMSM drive systems.
\end{abstract}


\IEEEpeerreviewmaketitle

\section{Introduction}
\IEEEPARstart{P}{ermanent} magnet synchronous motors (PMSMs) are widely employed in electric vehicles, industrial automation, servo drives, robotics, and renewable energy conversion systems because of their high efficiency, compact structure, and high torque density \cite{li2022fcsmpc_overview}. Industrial PMSM drives employ PI current regulators commonly because of
their simple structure, low computational burden, and well-established
implementation. However, more demanding control requirements, such as
multivariable coordination and explicit operating constraints, generally
require additional control and protection mechanisms \cite{Song2024PITuning, Cai2025MPCOverview}.

Model predictive control (MPC) has emerged as an attractive solution for high-performance PMSM drives \cite{vazquez2014mpc_review,li2022fcsmpc_overview,brosch2021rls_mpcc,wang2020fpga_ccsmpc,luo2020sixphase_mpc}. By explicitly using a discrete motor model, MPC predicts future system behavior and determines the control action by minimizing a predefined objective function. This provides several advantages, including fast transient response, direct treatment of multivariable coupling, systematic handling of voltage and current constraints, and the flexibility to incorporate multiple control objectives within a unified framework. However, these advantages come at the cost of considerable online computation. In finite-control-set MPC (FCS-MPC), candidate switching states or voltage vectors must be evaluated at every sampling instant, whereas continuous-control-set MPC (CCS-MPC) generally requires solving an optimization problem online. Such computational requirements become particularly restrictive in high-speed motor drives and resource-constrained platforms, especially when the sampling period is only several tens of microseconds \cite{wang2020fpga_ccsmpc,shen2026tiny_nn}.

Various approaches have therefore been investigated to improve the real-time feasibility of MPC. Reduced candidate sets, voltage-vector preselection, analytical voltage-vector synthesis, and reduced-dimension formulations have been developed to simplify the predictive optimization \cite{yan2016hybrid_control_set,zhang2018double_vectors,luo2020sixphase_mpc,brosch2023time_optimal_mpc,preindl2013pmsm_mptc,karamanakos2020fcsmpc_guidelines}. Dedicated hardware, such as FPGA-based implementations, has also been employed to accelerate online computation \cite{wang2020fpga_ccsmpc}. In addition, switching-frequency-oriented FCS-MPC strategies have been investigated to shape the output spectrum, facilitate filter design, and avoid the excitation of resonances in power converter applications \cite{aguirre2021switching_frequency}. Although these methods can significantly reduce execution time, an online search or optimization procedure is generally still retained, and its computational demand tends to increase with the prediction horizon and control complexity.

Neural network based control provides an alternative approach by shifting the computationally intensive optimization from online operation to offline learning \cite{dragicevic2019weighting,novak2021imitation,wang2022ann_mpc, jakobeit2023meta_rl,wan2025rl_predictive, drgona2024dpc_guarantees,li2026dpc_tpel}. Once trained, a neural policy can directly generate the control action through a forward inference, offering low and deterministic execution latency. Neural networks have therefore been used to assist predictive control, tune weighting factors, or directly approximate MPC policies \cite{dragicevic2019weighting,novak2021imitation,wang2022ann_mpc}. For example, deep neural networks have been trained using offline-generated MPC solutions to reproduce the corresponding state-to-control mapping for PMSM drives \cite{wang2022ann_mpc,novak2021imitation}. However, such supervised approaches require a sufficiently large set of optimal control labels obtained by repeatedly solving the original MPC problem, and their performance depends strongly on the quality and coverage of the labeled dataset \cite{drgona2024dpc_guarantees}. Reinforcement learning provides another label-free alternative by optimizing the control policy through interactions with the system according to a designed reward function \cite{jakobeit2023meta_rl,wan2025rl_predictive}. However, it generally requires extensive training interactions and careful reward design, while exploratory actions make direct training on physical power converters difficult \cite{li2026dpc_tpel}.

Differentiable predictive control (DPC) provides a different learning scheme \cite{drgona2024dpc_guarantees,li2026dpc_tpel}. Instead of using MPC as a supervisory controller to generate optimal control labels, DPC directly embeds the system model, predictive objective, and operational constraints into a differentiable computational graph. The neural control policy is optimized by backpropagating the predictive control objective through the model, enabling a label-free training process without precomputed optimal control actions. After offline training, the online optimization of MPC is replaced by a deterministic neural-network inference. Therefore, DPC preserves the model-based predictive structure of MPC while benefiting from the computational efficiency of neural networks. Its application to power electronic systems in \cite{li2026dpc_tpel, Li2026DPCC} further demonstrated the potential of realizing predictive control using compact neural architectures without an online optimization solver.

Nevertheless, a significant gap remains between demonstrating DPC on laboratory-scale power-electronic platforms and deploying it in industrial systems. Previous DPC studies have primarily focused on validating the control concept using laboratory converters and rapid-control prototyping platforms, where computational resources are relatively abundant \cite{li2026dpc_tpel}. In contrast, industrial PMSM drives rely on resource-constrained embedded processors and impose strict requirements on execution time, memory footprint, deterministic sampling, and integration with the complete drive-control architecture. Therefore, an important question is whether the predictive behavior of DPC can be retained using a lightweight neural policy that satisfies the stringent timing constraints of an industrial motor-drive controller. Moreover, industrial sensorless drives must accommodate practical operating procedures such as rotor alignment, low-speed startup, and transition to position- and speed-estimator-based operation \cite{wu2023optimized_pll,xu2024sensorless_dbpcc,wei2024ultralocalized_pmsm}. Therefore, demonstrating the industrial applicability of DPC requires not only reducing its computational burden to meet embedded real-time constraints, but also verifying that the learned controller can be seamlessly integrated into a complete industrial drive-control architecture.

Motivated by these challenges, this paper proposes a differentiable predictive current control (DPCC) strategy as a lightweight neural-network-based alternative to online-optimized MPC for PMSM drives. The PMSM system model and predictive current-tracking objective are embedded into an offline differentiable training framework, in which the neural controller generates a candidate voltage sequence and the differentiable motor model propagates the currents over the prediction horizon. The tracking loss is then evaluated using the predicted currents, and, since the entire closed prediction loop is differentiable, the controller parameters can be optimized directly through backpropagation. Consequently, the control policy is learned from the predictive-control objective itself without requiring precomputed optimal MPC actions as training labels. Once trained, the neural policy directly generates the control action through a single forward inference, thereby eliminating the need for iterative online optimization.

The proposed DPCC is implemented and experimentally evaluated on an industrial PMSM drive platform at Danfoss Drives within a complete sensorless drive-control architecture. This system-level validation enables the proposed controller to be assessed under realistic drive operating conditions. Building on the proposed control formulation and its real-time implementation, the main contributions of this work are summarized as follows:

\begin{itemize}
    \item A lightweight, label-free DPCC framework is developed for PMSM drives by embedding the motor model and predictive current-tracking objective into differentiable training. The learned policy replaces online optimization with compact neural-network inference without requiring MPC-generated labels.

    \item An analytical MPC implementation is developed as a computational benchmark for DPCC. The predictive optimization is expanded offline into a state-dependent control law, enabling real-time implementation and comparison of execution time and computational scalability.

    \item The proposed DPCC is implemented on an industrial PMSM platform within a complete sensorless drive architecture, including rotor alignment, low-speed startup, and back-EMF-based position and speed estimation.
\end{itemize}

The rest of this paper is organized as follows. Section~II presents the PMSM drive model and formulates the baseline predictive current control problem. Section~III introduces the proposed DPCC framework and its label-free training formulation. Section~IV describes the testing platform, the sensorless control architecture, and the DPCC and MPC deployment details. Section~V reports the experimental results under different operating conditions. Finally, Section~VI concludes the paper.

\section{Conventional MPC Formulation for PMSM Drives}

\subsection{PMSM Model in the Synchronous Reference Frame}

\begin{figure}[!t]
    \centering
    \includegraphics[width=\columnwidth]{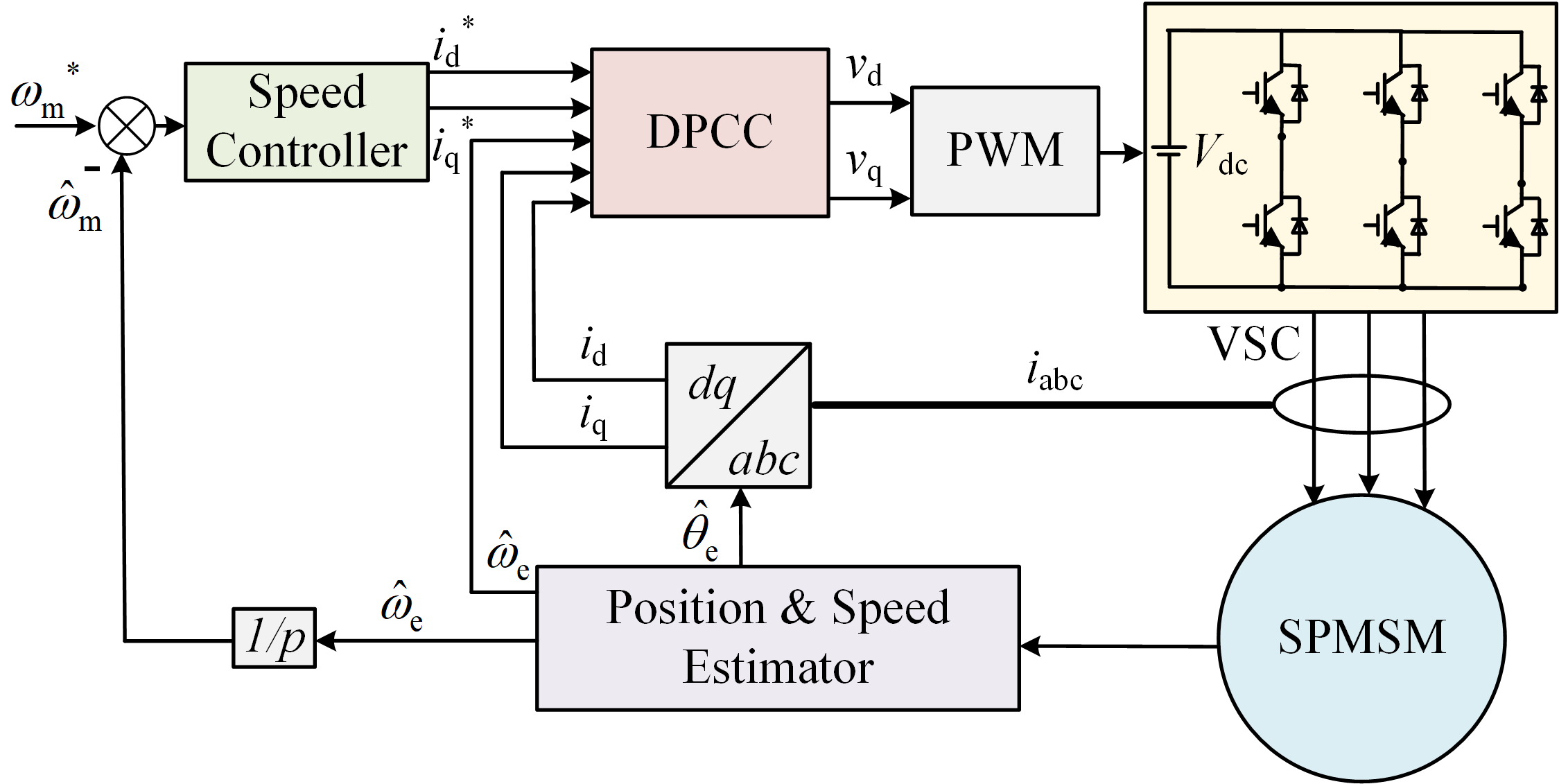}
    \caption{Overall control structure of the proposed sensorless PMSM drive with Differentiable predictive current control (DPCC).}
    \label{fig:control_diagram}
\end{figure}

The considered drive system consists of a surface-mounted permanent
magnet synchronous motor (SPMSM) supplied by a two-level voltage source
converter. Under field-oriented control (FOC), the three-phase stator
variables are transformed into the synchronous rotating $dq$ reference
frame using the rotor electrical angle $\theta_e$. The overall control diagram is shown in Fig.~\ref{fig:control_diagram}. In this reference frame,
the electrical dynamics of the PMSM are expressed as \cite{preindl2013pmsm_mptc}.
\begin{align}
\frac{di_d}{dt}
&=
\frac{1}{L_d}
\left(
v_d-R_s i_d+\omega_e L_q i_q
\right),
\label{eq:id_cont}
\\
\frac{di_q}{dt}
&=
\frac{1}{L_q}
\left(
v_q-R_s i_q-\omega_e L_d i_d-\omega_e\psi_f
\right),
\label{eq:iq_cont}
\end{align}
where $i_d$ and $i_q$ denote the $d$- and $q$-axis stator currents,
$v_d$ and $v_q$ are the corresponding voltage components, $R_s$ is
the stator resistance, $L_d$ and $L_q$ are the synchronous inductances,
$\psi_f$ is the permanent-magnet flux linkage, and $\omega_e$ denotes
the rotor electrical angular speed.

For control design, \eqref{eq:id_cont}--\eqref{eq:iq_cont} can be written
in the following form
\begin{equation}
\dot{\mathbf{i}}
=
\mathbf{A}_c(\omega_e)\mathbf{i}
+
\mathbf{B}_c\mathbf{v}
+
\mathbf{g}_c(\omega_e),
\label{eq:continuous_state_model}
\end{equation}
where
\begin{equation}
\mathbf{i}
=
\begin{bmatrix}
i_d\\i_q
\end{bmatrix},
\qquad
\mathbf{v}
=
\begin{bmatrix}
v_d\\v_q
\end{bmatrix},
\end{equation}
and
\begin{align}
\mathbf{A}_c(\omega_e)
&=
\begin{bmatrix}
-\dfrac{R_s}{L_d}
&
\dfrac{\omega_e L_q}{L_d}
\\[2mm]
-\dfrac{\omega_e L_d}{L_q}
&
-\dfrac{R_s}{L_q}
\end{bmatrix},
\\
\mathbf{B}_c
&=
\begin{bmatrix}
\dfrac{1}{L_d} & 0\\
0 & \dfrac{1}{L_q}
\end{bmatrix},
\\
\mathbf{g}_c(\omega_e)
&=
\begin{bmatrix}
0\\[1mm]
-\dfrac{\omega_e\psi_f}{L_q}
\end{bmatrix}.
\end{align}

The electromagnetic torque is given by
\begin{equation}
T_e=
\frac{3}{2}p
\left[
\psi_f i_q+
(L_d-L_q)i_di_q
\right],
\label{eq:torque}
\end{equation}
where $p$ is the number of pole pairs. For the SPMSM
considered in this work, $L_d \approx L_q$, and therefore
\begin{equation}
T_e \approx \frac{3}{2}p\psi_f i_q.
\label{eq:torque_spmsm}
\end{equation}
Accordingly, the electromagnetic torque is predominantly controlled
through the $q$-axis current.

The mechanical dynamics are described by
\begin{equation}
J\frac{d\omega_m}{dt}
=
T_e-T_L-B_v\omega_m,
\label{eq:mech}
\end{equation}
where $\omega_m$ is the mechanical angular speed in rad/s, $J$ is the combined
rotor and load inertia, $T_L$ is the load torque, and $B_v$ is the
viscous friction coefficient. The electrical
angular speed is then related to the mechanical speed by
\begin{equation}
\omega_e=p\omega_m.
\label{eq:speed_relation}
\end{equation}

\subsection{FOC-Based Drive Control Architecture}

The proposed DPCC and the benchmark MPC are both implemented within
the same cascaded FOC architecture. In the outer control loop, the error between the reference mechanical speed $\omega_m^\star$ and the actual or estimated speed $\omega_m$ is processed by the speed controller to generate the current
reference $i_q^\star$. For the surface-mounted PMSM operating below the
field-weakening region, the $d$-axis current reference is selected as
\begin{equation}
i_d^\star = 0,
\label{eq:id_ref}
\end{equation}
while $i_q^\star$ determines the required electromagnetic torque
according to \eqref{eq:torque_spmsm}. Equivalently, a torque reference
$T_e^\star$ corresponds to
\begin{equation}
i_q^\star
=
\frac{2T_e^\star}{3p\psi_f}.
\label{eq:iq_ref}
\end{equation}

The measured three-phase stator currents are transformed into the
synchronous reference frame using the electrical angle $\theta_e$.
The resulting currents $i_d$ and $i_q$, together with their references,
constitute the main feedback variables for the inner current controller.
The current controller generates the voltage references
$v_d^\star$ and $v_q^\star$, which are transformed back to the stationary
reference frame and applied to the VSC using pulse-width modulation.

\subsection{Prediction Model}

Since the electrical current dynamics are considerably faster than the
mechanical dynamics, the rotor electrical speed is assumed to remain
approximately constant over one current-control sampling interval
\cite{brosch2021rls_mpcc}.
Using forward-Euler discretization with sampling period $T_s$,
\eqref{eq:id_cont} and \eqref{eq:iq_cont} give
\begin{align}
i_d(k+1)
&=
\left(1-\frac{R_sT_s}{L_d}\right)i_d(k)
+\frac{\omega_e(k)L_qT_s}{L_d}i_q(k)
+\frac{T_s}{L_d}v_d(k),
\label{eq:id_discrete}
\\
i_q(k+1)
&=
-\frac{\omega_e(k)L_dT_s}{L_q}i_d(k)
+\left(1-\frac{R_sT_s}{L_q}\right)i_q(k)
\nonumber\\
&\quad
+\frac{T_s}{L_q}v_q(k)
-\frac{\omega_e(k)\psi_fT_s}{L_q}.
\label{eq:iq_discrete}
\end{align}
The discrete-time model can be expressed as
\begin{equation}
\mathbf{i}_{k+1}
=
\mathbf{A}_k\mathbf{i}_k
+
\mathbf{B}\mathbf{v}_k
+
\mathbf{g}_k,
\label{eq:discrete_model}
\end{equation}
where
\begin{equation}
\mathbf{A}_k
=
\begin{bmatrix}
1-\dfrac{R_sT_s}{L_d}
&
\dfrac{\omega_e(k)L_qT_s}{L_d}
\\[2mm]
-\dfrac{\omega_e(k)L_dT_s}{L_q}
&
1-\dfrac{R_sT_s}{L_q}
\end{bmatrix},
\end{equation}
\begin{equation}
\mathbf{B}
=
\begin{bmatrix}
\dfrac{T_s}{L_d} & 0\\
0 & \dfrac{T_s}{L_q}
\end{bmatrix},
\end{equation}
and
\begin{equation}
\mathbf{g}_k
=
\begin{bmatrix}
0\\[1mm]
-\dfrac{\omega_e(k)\psi_fT_s}{L_q}
\end{bmatrix}.
\end{equation}

\subsection{Predictive Current Control Formulation}

Based on \eqref{eq:discrete_model}, the CCS-MPC predicts the evolution of the stator currents over a prediction
horizon $N_p$. In this work, a five-step prediction horizon,
i.e., $N_p=5$, is adopted for both MPC and DPCC. This horizon was selected as a compromise between retaining short-term predictive behavior and keeping the training-time output dimension small enough for a deployed network. A longer horizon may further exploit the predictive capability of DPC, but it also increases the number of training-time voltage outputs and the complexity of the embedded benchmark MPC implementation \cite{li2026dpc_tpel}.

\begin{equation}
\mathbf{V}_k=
\left[
\mathbf{v}_k^\top,
\mathbf{v}_{k+1}^\top,
\ldots,
\mathbf{v}_{k+N_p-1}^\top
\right]^\top .
\label{eq:mpc_voltage_sequence}
\end{equation}

Starting from the measured current state $\mathbf{i}_k$, the future
currents are recursively predicted using \eqref{eq:discrete_model}.
The objective of the current controller is to force the predicted
$d$- and $q$-axis currents to follow their corresponding references.
Accordingly, the current-tracking cost is defined as \cite{karamanakos2020fcsmpc_guidelines}

\begin{equation}
J_{\mathrm{MPC}}
=
\sum_{j=1}^{N_p}
\left\|
\mathbf{i}_{k+j}
-
\mathbf{i}_{k+j}^{\star}
\right\|_{\mathbf{Q}}^2 ,
\label{eq:mpc_tracking_cost}
\end{equation}

where

\begin{equation}
\mathbf{i}_{k+j}^{\star}
=
\begin{bmatrix}
i_d^{\star}(k+j)\\
i_q^{\star}(k+j)
\end{bmatrix},
\qquad
\mathbf{Q}
=
\begin{bmatrix}
q_d & 0\\
0 & q_q
\end{bmatrix},
\end{equation}
and $q_d$ and $q_q$ determine the weights of the two
current-tracking errors.

{\color{black}

The MPC problem is therefore formulated as
\begin{align}
\mathbf{V}_k^\star
=
\arg\min_{\mathbf{V}_k}
\quad&
\sum_{j=1}^{N_p}
\left\|
\mathbf{i}_{k+j}
-
\mathbf{i}_{k+j}^{\star}
\right\|_{\mathbf{Q}}^2
\label{eq:mpc_problem}
\\
\mathit{s.t.}\quad
&
\mathbf{i}_{k+j+1}
=
\mathbf{A}_{k+j}\mathbf{i}_{k+j}
+
\mathbf{B}\mathbf{v}_{k+j}
+
\mathbf{g}_{k+j},
\nonumber\\
&
\|\mathbf{v}_{k+j}\|_2\leq V_{\max},
\quad
\nonumber\\
&
\|\mathbf{i}_{k+j+1}\|_2\leq I_{\max},
\quad 
\nonumber\\
&
j=0,\ldots,N_p-1.
\nonumber
\end{align}

{\color{black}
Here, $V_{\max}$ and $I_{\max}$ denote the voltage
and current limits, respectively. 
}

}

After solving \eqref{eq:mpc_problem}, only the first voltage vector of
the optimal sequence is applied to the inverter,

\begin{equation}
\mathbf{v}_k^\star
=
\begin{bmatrix}
v_d^\star(k)\\
v_q^\star(k)
\end{bmatrix}.
\label{eq:first_mpc_action}
\end{equation}

At the next sampling instant, the newly measured currents and rotor
speed are used to update the prediction model, and the optimization is
solved again. This receding horizon procedure allows MPC to continuously
adapt the voltage command to the instantaneous operating condition.
However, the prediction and optimization must be repeated within every
current-control period resulting in heavy online computation.

\section{Proposed Differentiable Predictive Current Control}

The conventional MPC described in the previous section determines the
optimal control sequence by repeatedly solving an optimization problem. The proposed DPCC follows the same model-based predictive principle but transfers the optimization process from online operation to offline learning. Specifically, the PMSM prediction model is embedded into a differentiable computational graph, through which the parameters of a neural control policy are directly
optimized according to the predicted current-tracking performance.
After training, the control policy generates the voltage control action
through a single forward inference without solving the predictive
optimization problem online.

For consistency with the MPC benchmark, a prediction horizon of $N_p=5$ is adopted throughout this work. Since the prediction interval covers only five current-control sampling periods, the current references and rotor electrical speed are assumed to remain approximately constant within one prediction horizon, i.e.,
\begin{equation}
\mathbf{i}_{k+j|k}^{\star}
\approx
\mathbf{i}_{k}^{\star},
\qquad
\omega_{e,k+j|k}
\approx
\omega_e(k),
\quad
j=0,\ldots,N_p-1.
\label{eq:dpcc_horizon_assumption}
\end{equation}

\subsection{Differentiable Control Policy}

The input of the DPCC policy consists of the measured $dq$-axis currents,
their references, and the electrical rotor speed,
\begin{equation}
\mathbf{x}_k
=
\begin{bmatrix}
i_d(k) &
i_q(k) &
i_d^{\star}(k) &
i_q^{\star}(k) &
\omega_e(k)
\end{bmatrix}^{\top}.
\label{eq:dpcc_input}
\end{equation}
During offline training, the neural policy produces a sequence of
control actions over the complete prediction horizon,
\begin{equation}
\mathbf{V}_{\theta,k}
=
f_{\theta}(\mathbf{x}_k)
=
\begin{bmatrix}
\mathbf{v}_{k|k}^{\top} &
\mathbf{v}_{k+1|k}^{\top} &
\cdots &
\mathbf{v}_{k+N_p-1|k}^{\top}
\end{bmatrix}^{\top},
\label{eq:dpcc_policy}
\end{equation}
where
\begin{equation}
\mathbf{v}_{k+j|k}
=
\begin{bmatrix}
v_{d,k+j|k} &
v_{q,k+j|k}
\end{bmatrix}^{\top}.
\label{eq:dpcc_voltage_pair}
\end{equation}
Therefore, for $N_p=5$, the training-time policy contains ten output variables, corresponding to five predicted pairs of $d$- and $q$-axis control actions.

A lightweight feedforward neural network is adopted to enable real-time implementation on an industrial DSP. To balance control performance and computational complexity, a pruning-based architecture reduction procedure is applied to remove redundant neurons while preserving current-tracking performance. The final network consists of two fully connected hidden layers with six neurons each. With five input variables and ten outputs representing the five-step voltage sequence during training, its architecture is $5$--$6$--$6$--$10$. The forward propagation during training is expressed as

\begin{align}
\mathbf{h}_1
&=
\sigma
\left(
\mathbf{W}_1\mathbf{x}_k+\mathbf{b}_1
\right),
\label{eq:dpcc_layer1}
\\
\mathbf{h}_2
&=
\sigma
\left(
\mathbf{W}_2\mathbf{h}_1+\mathbf{b}_2
\right),
\label{eq:dpcc_layer2}
\\
\mathbf{V}_{\theta,k}
&=
\mathbf{W}_3\mathbf{h}_2+\mathbf{b}_3,
\label{eq:dpcc_layer3}
\end{align}
where $\sigma(\cdot)$ denotes the ReLU activation function. The dimensions of the trainable matrices are
$\mathbf{W}_1\in\mathbb{R}^{6\times5}$,
$\mathbf{W}_2\in\mathbb{R}^{6\times6}$, and
$\mathbf{W}_3\in\mathbb{R}^{10\times6}$, resulting in only 148
trainable weights and biases despite explicitly representing a
five-step control sequence.

It should be noted that only the first voltage vector is required during real-time operation, consistent with the receding-horizon principle of MPC. Therefore, after training, only the first two outputs corresponding to $v_d^{*}(k)$ and $v_q^{*}(k)$ are retained for online inference, resulting in an effective $5$--$6$--$6$--$2$ deployment architecture. The deployed policy consequently contains only 92 active weights and biases, while its parameters are optimized using the full five-step predictive objective during offline training.

\subsection{DPC Training}

\begin{figure*}[!t]
    \centering
    \includegraphics[width=1\textwidth]{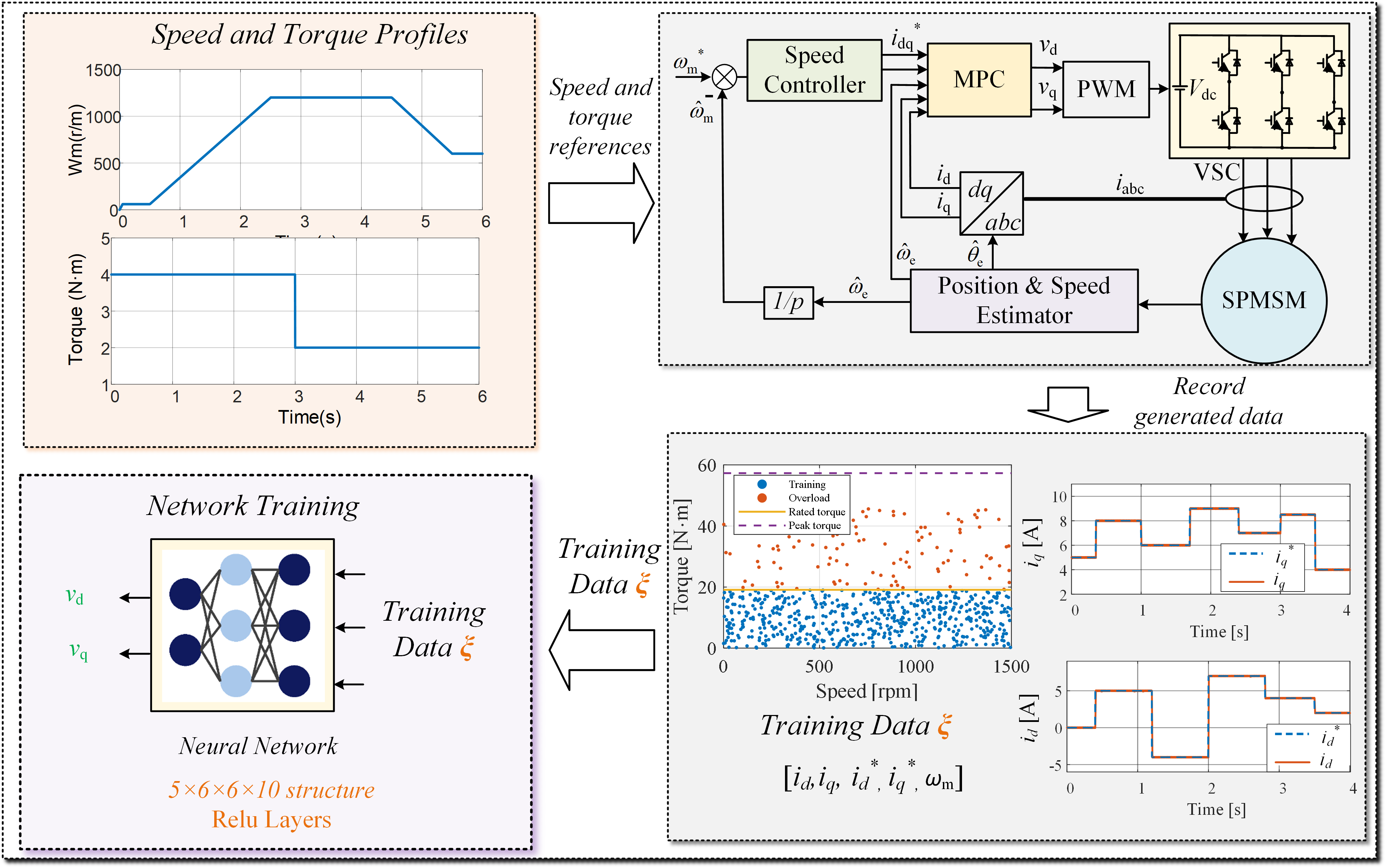}
    \caption{Label-free offline training workflow for the proposed DPCC. Operating samples provide current states, current references, and speed information, while the neural policy is optimized through the differentiable PMSM prediction model and predictive tracking loss without MPC-generated voltage labels.}
    \label{fig:training_diagram}
\end{figure*}

Unlike supervised MPC approximation, DPCC does not require optimal voltage labels. The sampled operating points provide only initial conditions and reference commands, while the policy is optimized directly through the differentiable PMSM rollout and predictive loss, as illustrated in Fig.~\ref{fig:training_diagram}.

For each training sample, the measured current state is
used as the initial condition,
\begin{equation}
\hat{\mathbf{i}}_{k|k}
=
\mathbf{i}_k.
\label{eq:dpcc_initial_state}
\end{equation}
The ten network outputs in \eqref{eq:dpcc_policy} are divided into five
consecutive voltage vectors and applied sequentially to the discrete
PMSM prediction model. The predicted current trajectory is obtained from
\begin{equation}
\hat{\mathbf{i}}_{k+j+1|k}
=
\mathbf{A}_{k+j}
\hat{\mathbf{i}}_{k+j|k}
+
\mathbf{B}
\mathbf{v}_{k+j|k}
+
\mathbf{g}_{k+j},
\quad
j=0,\ldots,N_p-1.
\label{eq:dpcc_rollout}
\end{equation}
For the five-step prediction considered in this work,
\eqref{eq:dpcc_rollout} generates
$\hat{\mathbf{i}}_{k+1|k},\ldots,
\hat{\mathbf{i}}_{k+5|k}$ from the voltage sequence produced by the
neural policy.

The quality of this sequence is evaluated according to the
$d$- and $q$-axis current-tracking errors. The baseline multi-step predictive
loss is defined as
\begin{equation}
\mathcal{L}_k(\theta)
=
\sum_{j=1}^{N_p}
\left\|
\hat{\mathbf{i}}_{k+j|k}
-
\mathbf{i}_{k+j|k}^{\star}
\right\|_{\mathbf{Q}}^2,
\label{eq:dpcc_loss}
\end{equation}
where
\begin{equation}
\mathbf{Q}
=
\begin{bmatrix}
q_d & 0\\
0 & q_q
\end{bmatrix}.
\end{equation}
Equivalently, the loss can be written explicitly as
\begin{align}
\mathcal{L}_k(\theta)
=
\sum_{j=1}^{N_p}
\Big[
&
q_d
\left(
\hat{i}_{d,k+j|k}
-
i_{d,k+j|k}^{\star}
\right)^2
\nonumber\\
+
&
q_q
\left(
\hat{i}_{q,k+j|k}
-
i_{q,k+j|k}^{\star}
\right)^2
\Big].
\label{eq:dpcc_loss_expanded}
\end{align}

The network parameters are then obtained by minimizing the expected
predictive loss over the training dataset $\mathcal{D}$,
\begin{equation}
\theta^{\star}
=
\arg\min_{\theta}
\;
\mathbb{E}_{\mathbf{x}_k\sim\mathcal{D}}
\left[
\mathcal{L}_k(\theta)
\right].
\label{eq:dpcc_training}
\end{equation}
Since the PMSM prediction model and the neural-network operations are
differentiable with respect to $\theta$, the gradient of
\eqref{eq:dpcc_training} can be propagated backward through the entire
five-step prediction process. Consequently, the network parameters are
updated according to the current trajectories produced by the control
policy itself.

\subsection{Training Data Generation and Network Optimization}

To improve the operating range coverage of the learned controller,
training trajectories are generated under different rotor speeds,
load-torque levels, and reference transitions. The measured current states $i_d$ and $i_q$ are sampled within
$[-15,15]$~A to cover both normal operation and transient current
excursions. Although the considered surface-mounted PMSM normally
operates with $i_d^\star=0$ below the field-weakening region, a wider
range of $i_d^\star$ is included in the training set to accommodate
the direct $d$-axis current-reference step tests used for evaluating
the inner current-control dynamics. Accordingly, $i_d^\star$ is sampled
within $[-10,10]$~A. The torque-producing $q$-axis current reference
$i_q^\star$ is sampled within $[-12,12]$~A to cover both motoring and
regenerative operating conditions. The electrical angular speed is
varied from 0 to 650~rad/s, covering the complete rated-speed range of
the four-pole-pair, 1500-r/min experimental motor.

The resulting dataset is represented by
$\mathbf{X}\in\mathbb{R}^{5\times50000}$. The dataset is
randomly divided into 80\% for training and 20\% for validation,
corresponding to 40,000 and 10,000 samples, respectively. During each optimization iteration, a batch of operating samples is
selected from $\mathcal{D}$. The neural network first produces the
five-step voltage sequence in \eqref{eq:dpcc_policy}. The discrete PMSM
model is then recursively evaluated according to
\eqref{eq:dpcc_rollout}, after which the corresponding multi-step
current-tracking loss is obtained from \eqref{eq:dpcc_loss}. The loss is
backpropagated through the prediction model and neural network to update
the policy parameters.

\begin{figure}[!t]
    \centering
    \includegraphics[width=\columnwidth]{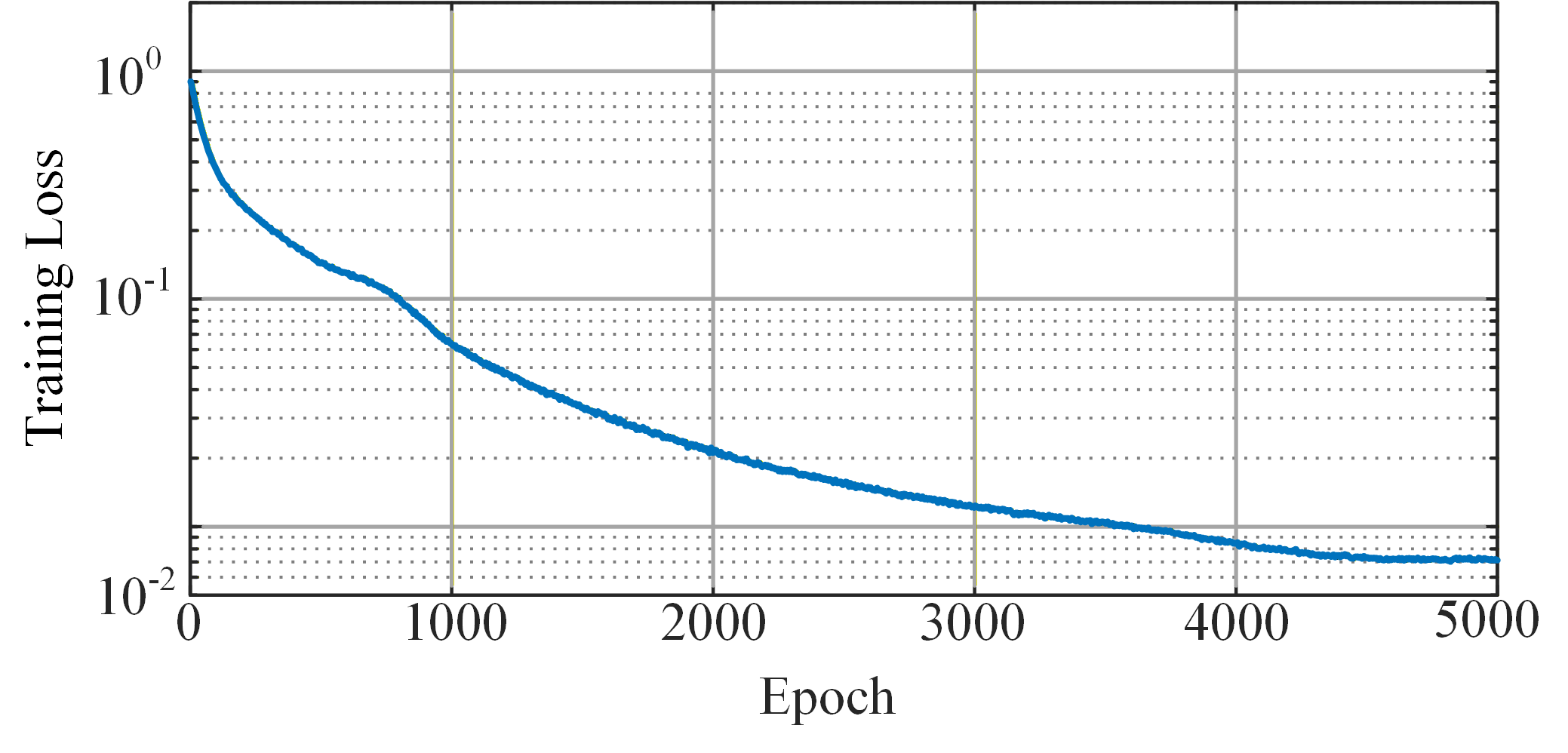}
    \caption{Training loss histories during offline
    optimization of the proposed DPCC policy.}
    \label{fig:training_loss}
\end{figure}

The Adam optimizer is employed with a learning rate of $10^{-3}$ and a
mini-batch size of 256. The evolution of the training loss during network optimization is shown in Fig.~\ref{fig:training_loss}, providing a direct assessment of the
convergence behavior.

After convergence, the optimized network parameters
\begin{equation}
\theta^\star
=
\left\{
\mathbf{W}_1^\star,\mathbf{b}_1^\star,
\mathbf{W}_2^\star,\mathbf{b}_2^\star,
\mathbf{W}_3^\star,\mathbf{b}_3^\star
\right\}
\end{equation}
are fixed for real-time implementation.

\section{Embedded Implementation and Sensorless Drive Operation}

The practical implementation is examined from three aspects. First, the MPC benchmark is reformulated into an offline-expanded analytical implementation for real-time comparison with DPCC. Second, the computational scalability of MPC and DPCC is analyzed for more general predictive-control formulations. Third, a sensorless startup and speed-estimation scheme is integrated into the drive for experimental evaluation without mechanical position feedback.

\subsection{Offline-Expanded Analytical Implementation of the MPC Benchmark}
\label{sec:symbolic_mpc}

A direct real-time implementation of conventional MPC requires the prediction matrices to be constructed and the corresponding optimization problem to be solved at every sampling instant. Although such an implementation follows the MPC formulation directly, its execution time exceeds the available sampling interval on the target embedded processor, preventing direct online implementation in the considered PMSM drive. To establish a feasible real-time benchmark, the online MPC computations are first expanded offline into explicit algebraic expressions prior to deployment, and these expressions are then evaluated directly on the embedded processor at each sampling instant. This offline-expanded analytical implementation eliminates repeated online matrix construction and substantially reduces the computational overhead while preserving the parameter dependence of the original MPC formulation. 

By recursively applying the one-step discrete-time model in
\eqref{eq:discrete_model}, the predicted current trajectory over
$N_p$ steps can be written in the stacked form

\begin{equation}
\mathbf{I}_k
=
\mathbf{F}_k\mathbf{i}_k
+
\mathbf{G}_k\mathbf{V}_k
+
\mathbf{c}_k,
\label{eq:mpc_stacked_prediction}
\end{equation}
where

\begin{equation}
\mathbf{I}_k
=
\begin{bmatrix}
\mathbf{i}_{k+1|k}^{\top} &
\mathbf{i}_{k+2|k}^{\top} &
\cdots &
\mathbf{i}_{k+N_p|k}^{\top}
\end{bmatrix}^{\top},
\end{equation}
and
\begin{equation}
\mathbf{V}_k
=
\begin{bmatrix}
\mathbf{v}_{k|k}^{\top} &
\mathbf{v}_{k+1|k}^{\top} &
\cdots &
\mathbf{v}_{k+N_p-1|k}^{\top}
\end{bmatrix}^{\top}.
\end{equation}
To express the matrices explicitly, define
the state-transition product
\begin{equation}
\boldsymbol{\Phi}_{r,s}
=
\begin{cases}
\mathbf{A}_{k+r-1}\mathbf{A}_{k+r-2}\cdots\mathbf{A}_{k+s},
& r>s,\\
\mathbf{I}_2, & r=s,
\end{cases}
\end{equation}
for $r=1,\ldots,N_p$ and $s=0,\ldots,r$. Then
\begin{equation}
\mathbf{F}_k
=
\begin{bmatrix}
\boldsymbol{\Phi}_{1,0}\\
\boldsymbol{\Phi}_{2,0}\\
\vdots\\
\boldsymbol{\Phi}_{N_p,0}
\end{bmatrix},
\end{equation}
\begin{equation}
\mathbf{G}_k
=
\begin{bmatrix}
\mathbf{B} & \mathbf{0} & \cdots & \mathbf{0}\\
\boldsymbol{\Phi}_{2,1}\mathbf{B} & \mathbf{B} & \cdots & \mathbf{0}\\
\vdots & \vdots & \ddots & \vdots\\
\boldsymbol{\Phi}_{N_p,1}\mathbf{B} &
\boldsymbol{\Phi}_{N_p,2}\mathbf{B} &
\cdots &
\mathbf{B}
\end{bmatrix},
\end{equation}
and
\begin{equation}
\mathbf{c}_k
=
\begin{bmatrix}
\mathbf{g}_{k}\\
\boldsymbol{\Phi}_{2,1}\mathbf{g}_{k}+\mathbf{g}_{k+1}\\
\vdots\\
\displaystyle
\sum_{s=0}^{N_p-1}
\boldsymbol{\Phi}_{N_p,s+1}\mathbf{g}_{k+s}
\end{bmatrix}.
\end{equation}
Therefore, $\mathbf{F}_k$, $\mathbf{G}_k$, and $\mathbf{c}_k$ are
constructed directly from the one-step matrices $\mathbf{A}_k$,
$\mathbf{B}$, and $\mathbf{g}_k$ in \eqref{eq:discrete_model}.

Substituting \eqref{eq:mpc_stacked_prediction} into the current-tracking
objective gives the quadratic form

\begin{equation}
J_{\mathrm{MPC}}
=
\frac{1}{2}
\mathbf{V}_k^{\top}
\mathbf{H}_k
\mathbf{V}_k
-
\mathbf{b}_k^{\top}
\mathbf{V}_k
+
C,
\label{eq:mpc_quadratic_form}
\end{equation}
where $\mathbf{H}_k$ is the Hessian matrix, $\mathbf{b}_k$ contains the
dependence on the measured currents, current references, and operating
conditions, and $C$ is independent of the control sequence. The
stationary condition of \eqref{eq:mpc_quadratic_form} leads to

\begin{equation}
\mathbf{H}_k\mathbf{V}_k^{\star}
=
\mathbf{b}_k.
\label{eq:mpc_linear_system}
\end{equation}

Rather than constructing $\mathbf{H}_k$ and solving
\eqref{eq:mpc_linear_system} using a general matrix inversion
during every control cycle, the complete calculation is algebraically expanded offline prior to deployment. In particular, the symmetric
positive-definite matrix $\mathbf{H}_k$ is factorized using the classical
$LDL^{\top}$ decomposition \cite{martin1965symmetric_decomposition} as

\begin{equation}
\mathbf{H}_k
=
\mathbf{L}_k
\mathbf{D}_k
\mathbf{L}_k^{\top},
\label{eq:ldlt}
\end{equation}
where $\mathbf{L}_k$ is a unit lower-triangular matrix and
$\mathbf{D}_k$ is diagonal. The solution of
\eqref{eq:mpc_linear_system} is then obtained through

\begin{align}
\mathbf{L}_k\mathbf{y}_k
&=
\mathbf{b}_k,
\\
\mathbf{D}_k\mathbf{z}_k
&=
\mathbf{y}_k,
\\
\mathbf{L}_k^{\top}\mathbf{V}_k^{\star}
&=
\mathbf{z}_k.
\label{eq:ldlt_solution}
\end{align}

The entries involved in the factorization and the forward/backward
substitutions are derived as closed-form scalar expressions and subsequently expanded into
scalar arithmetic expressions. Consequently, the embedded controller
does not perform a generic matrix inversion or invoke an optimization
solver during real-time operation. It directly evaluates the derived expressions using the current measured states and
references.

It should be noted that this procedure differs from the conventional offline matrix
precomputation. Rather than storing a fixed gain matrix for one
operating condition, quantities that vary during operation, such as
the electrical speed and current references, are retained as variables
during the derivation. The resulting control law can
therefore be expressed in the general form

\begin{equation}
\mathbf{V}_k^{\star}
=
\mathcal{S}
\left(
\mathbf{i}_k,
\mathbf{i}_k^{\star},
\omega_e(k)
\right),
\label{eq:symbolic_mpc_solution}
\end{equation}
where $\mathcal{S}(\cdot)$ denotes the offline-expanded analytical MPC
solution. The same derivation procedure can be regenerated when the
prediction horizon or model formulation is changed, which makes the
approach more general than simply precomputing a fixed set of matrices. Consistent with the receding-horizon principle, only the first voltage
vector of \eqref{eq:symbolic_mpc_solution} is applied to the inverter. The offline-expanded implementation therefore represents a computationally optimized MPC baseline. 

\subsection{Computational Complexity and Constrained-Control Evaluation}
\label{sec:complexity_scalability}
\begingroup\color{black}

The offline-expanded analytical MPC in Section~\ref{sec:symbolic_mpc} provides an efficient implementation for the formulation used in the experiments. Its reduced online burden allows MPC and DPCC to be implemented on the same embedded platform and compared under identical operating conditions. The same analytical procedure can also be regenerated for different prediction horizons.

To examine the computational scalability beyond this specific implementation, consider a more general predictive-control formulation. For a prediction horizon $N_p$, the stacked voltage sequence is
\begin{equation}
\mathbf{U}=\mathbf{V}_k=
\begin{bmatrix}
v_{d,k|k} & v_{q,k|k} & \cdots & v_{d,k+N_p-1|k} & v_{q,k+N_p-1|k}
\end{bmatrix}^{\top},
\end{equation}
giving
\begin{equation}
n_u=2N_p
\end{equation}
decision variables.

\blueadd{For the constrained formulation in \eqref{eq:mpc_problem},} the circular voltage and current feasible regions can be approximated by inscribed polygons to obtain a standard QP formulation. Let $N_f$ denote the number of polygon facets. The constraints can be expressed as
\begin{equation}
\mathbf{a}_{\ell}^{\top}\mathbf{v}_j
\leq
V_{\max}\cos\left(\frac{\pi}{N_f}\right),
\qquad
\ell=1,\ldots,N_f,
\end{equation}
and
\begin{equation}
\mathbf{a}_{\ell}^{\top}\mathbf{i}_{j|k}
\leq
I_{\max}\cos\left(\frac{\pi}{N_f}\right),
\qquad
\ell=1,\ldots,N_f,
\end{equation}
where
\begin{equation}
\mathbf{a}_{\ell}
=
\begin{bmatrix}
\cos(2\pi\ell/N_f) &
\sin(2\pi\ell/N_f)
\end{bmatrix}^{\top}.
\end{equation}
A larger $N_f$ gives a tighter approximation but introduces more inequalities. Considering both voltage and current limits, the number of constraints scales approximately as
\begin{equation}
m=2N_fN_p.
\end{equation}
For a dense factorization-based QP implementation, the corresponding solution effort can be characterized by
\begin{equation}
\mathcal{C}_{\mathrm{MPC}}
=
\mathcal{O}\left((n_u+m)^3\right).
\label{eq:cmpc_complexity}
\end{equation}
This expression illustrates the scaling trend rather than the exact cost of a particular solver. Sparse or first-order solvers can reduce the practical computation \cite{wang2010fast_mpc,richter2014accelerated_mpc}, but the online workload still grows with the problem dimension.

In contrast, DPCC requires only a feedforward neural-network inference online. For the deployed $5$--$w$--$w$--$2$ architecture, the number of weights and biases is
\begin{equation}
N_{\theta}
=
(5w+w)+(w^2+w)+(2w+2)
=
w^2+9w+2,
\label{eq:dpc_params}
\end{equation}
and one forward inference requires
\begin{equation}
\mathcal{C}_{\mathrm{DPCC}}
=
5w+w^2+2w
=
w^2+7w
\label{eq:dpc_complexity}
\end{equation}
multiply--accumulate operations. Thus, increasing $N_p$ directly enlarges the online MPC problem, whereas for DPCC it mainly affects the offline predictive objective and, indirectly, the network capacity required to represent the learned policy. The online controller itself remains a deterministic forward pass.

{\color{black}
To further examine this difference, a supplementary constrained-control
study considers the voltage and predicted-current limits introduced in
\eqref{eq:mpc_problem}, with
\begin{equation}
V_{\max}=325~\mathrm{V},\qquad I_{\max}=10~\mathrm{A}.
\label{eq:supp_constraints}
\end{equation}
}

The evaluation uses 8000 training samples, 2000 independent validation
samples, and 3000 out-of-range samples with the electrical-speed range
extended by 20\%.

\begin{figure}[!t]
    \centering
    \IfFileExists{violation.png}{%
        \includegraphics[width=0.8\columnwidth]{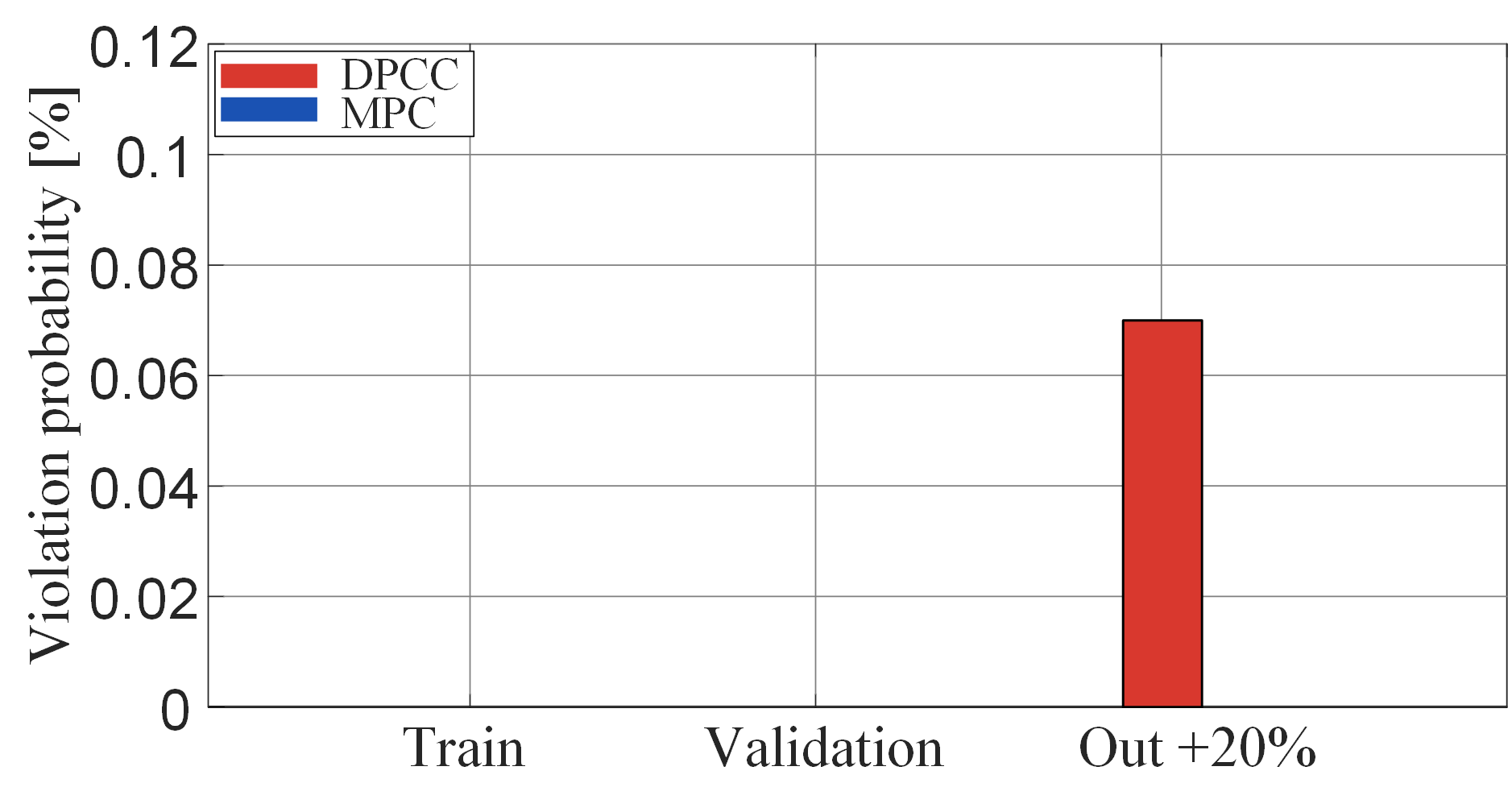}%
    }{%
        \figplaceholder{35mm}{\columnwidth}%
    }
    \caption{Constraint-violation probability of \blueadd{DPCC and MPC} over the supplementary evaluation sets.}
    \label{fig:constraint_violation_summary}
\end{figure}

\blueadd{For each evaluation set, the constraint-violation probability is
$P_{\mathrm{vio}}=100N_{\mathrm{vio}}/N_{\mathrm{total}}\%$, where
$N_{\mathrm{vio}}$ counts samples that violate either the voltage
or current limit (counted only once if both limits are exceeded).
The criterion is applied consistently to the evaluated control action
and its resulting current prediction.}
As shown in Fig.~\ref{fig:constraint_violation_summary}, no violations
are observed for the \blueadd{DPCC} training or validation samples,
while the violation probability is approximately 0.07\% under the 20\%
speed extension.

\begin{figure*}[!t]
    \centering
    \IfFileExists{Outputs.png}{%
        \includegraphics[width=0.9\textwidth]{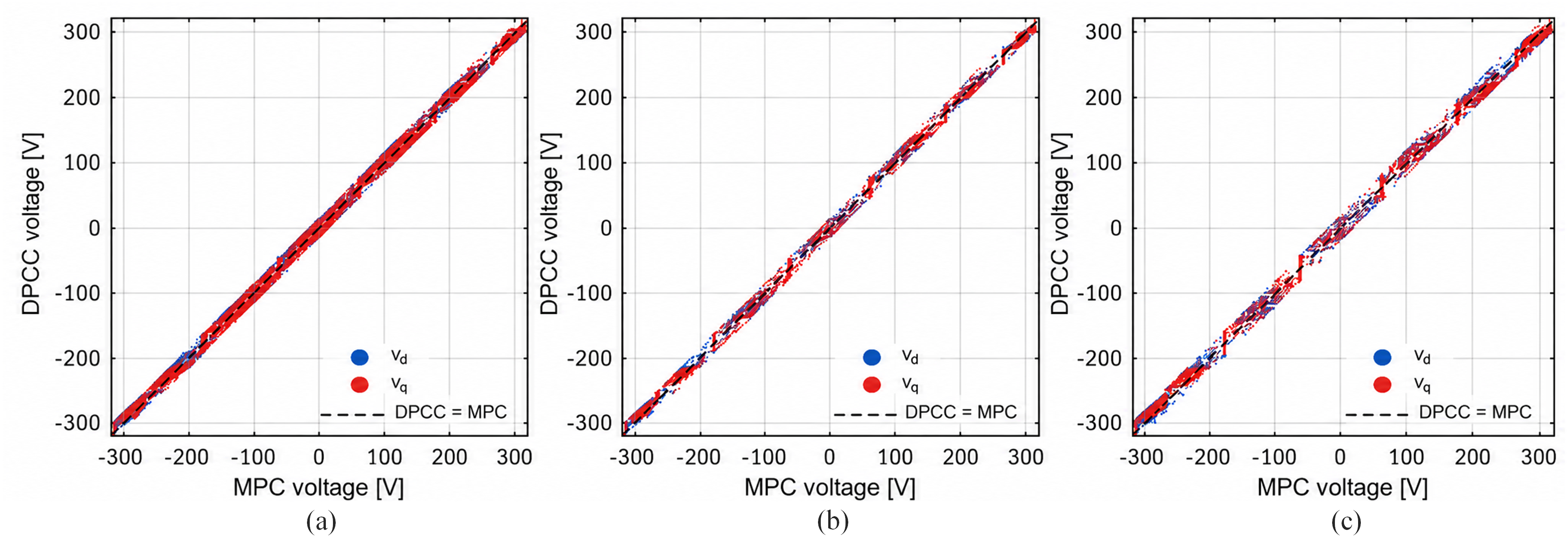}%
    }{%
        \figplaceholder{42mm}{0.9\textwidth}%
    }
    \caption{Comparison of the first voltage action generated by \blueadd{DPCC and MPC}: (a) training region with 8000 samples; (b) validation region with 2000 samples; and (c) out-of-range operation with 3000 samples and a 20\% extension of the electrical-speed range.}
    \label{fig:constrained_control_actions}
\end{figure*}

Fig.~\ref{fig:constrained_control_actions} shows close agreement between the \blueadd{DPCC and MPC} voltage actions in both the training and validation regions. A slightly larger dispersion appears under the 20\% speed extension, while the overall mapping remains close to the \blueadd{MPC} solution. \blueadd{The mean Euclidean voltage errors relative to MPC,
normalized by $V_{\max}$, are approximately 4.5\%, 4.1\%,
and 7.5\% for the training, validation, and out-of-range sets,
respectively.}

Fig. \ref{fig:constrained_boundary_response} further compares the current and voltage responses of MPC and DPCC under operation near the imposed constraints. Both controllers track the step changes in the current reference, while the voltage magnitude approaches the specified limit of 325 V during the transients. The current responses remain close to the imposed 10 A boundary, demonstrating comparable tracking behavior under constrained operating conditions.

\begin{figure}[!t]
    \centering
    \begin{minipage}[t]{0.49\columnwidth}
        \centering
        \includegraphics[width=\linewidth]{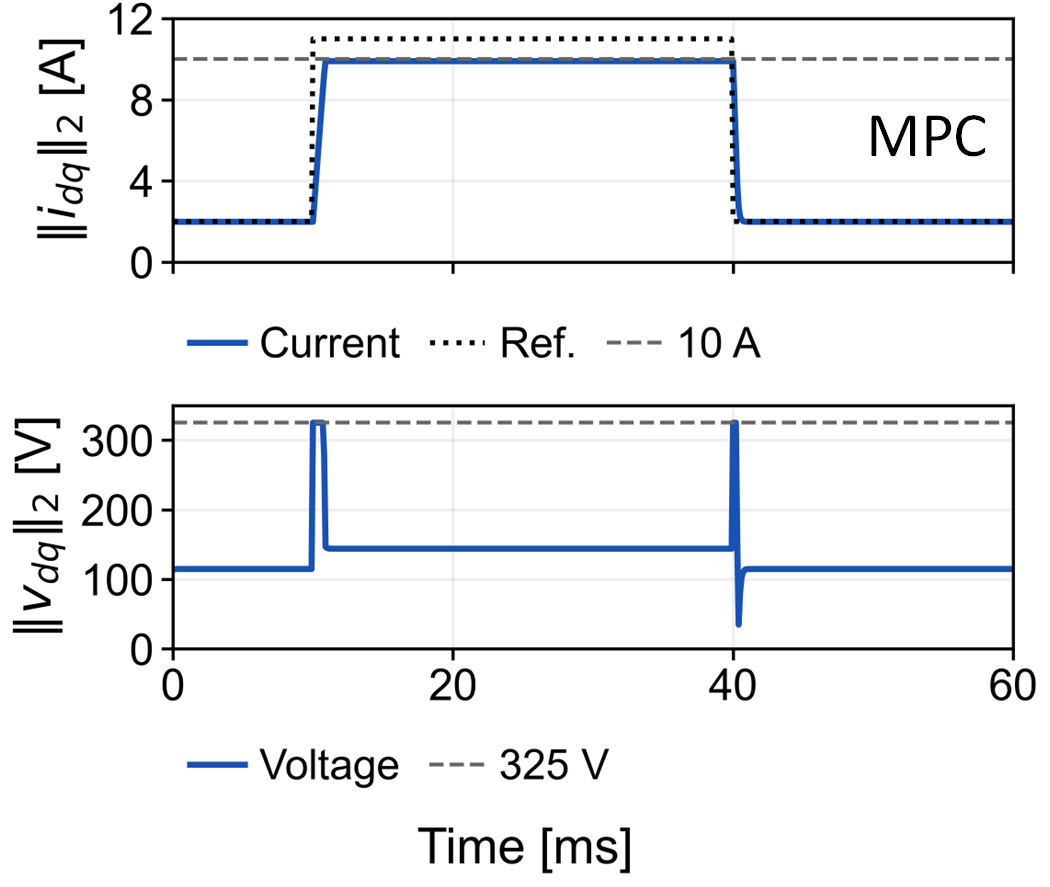}
        \centerline{(a)}
    \end{minipage}\hfill
    \begin{minipage}[t]{0.49\columnwidth}
        \centering
        \includegraphics[width=\linewidth]{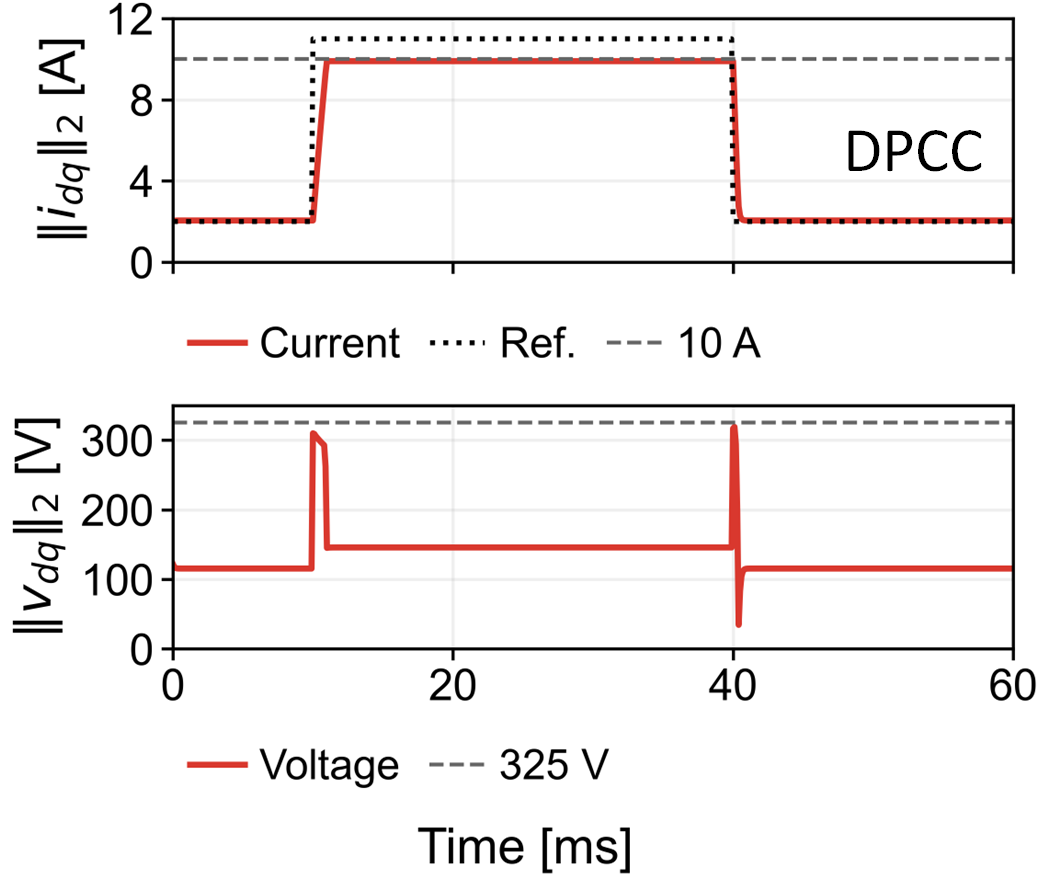}
        \centerline{(b)}
    \end{minipage}
    \caption{Current and voltage responses under
    operation near the imposed constraints:
    (a) MPC and (b) DPCC.
    The dashed lines indicate the current
    reference and the corresponding
    constraint limits.}
    \label{fig:constrained_boundary_response}
\end{figure}

\begin{figure}[!t]
    \centering
    \IfFileExists{Computationtimeconstrained.png}{%
        \includegraphics[width=0.8\columnwidth]{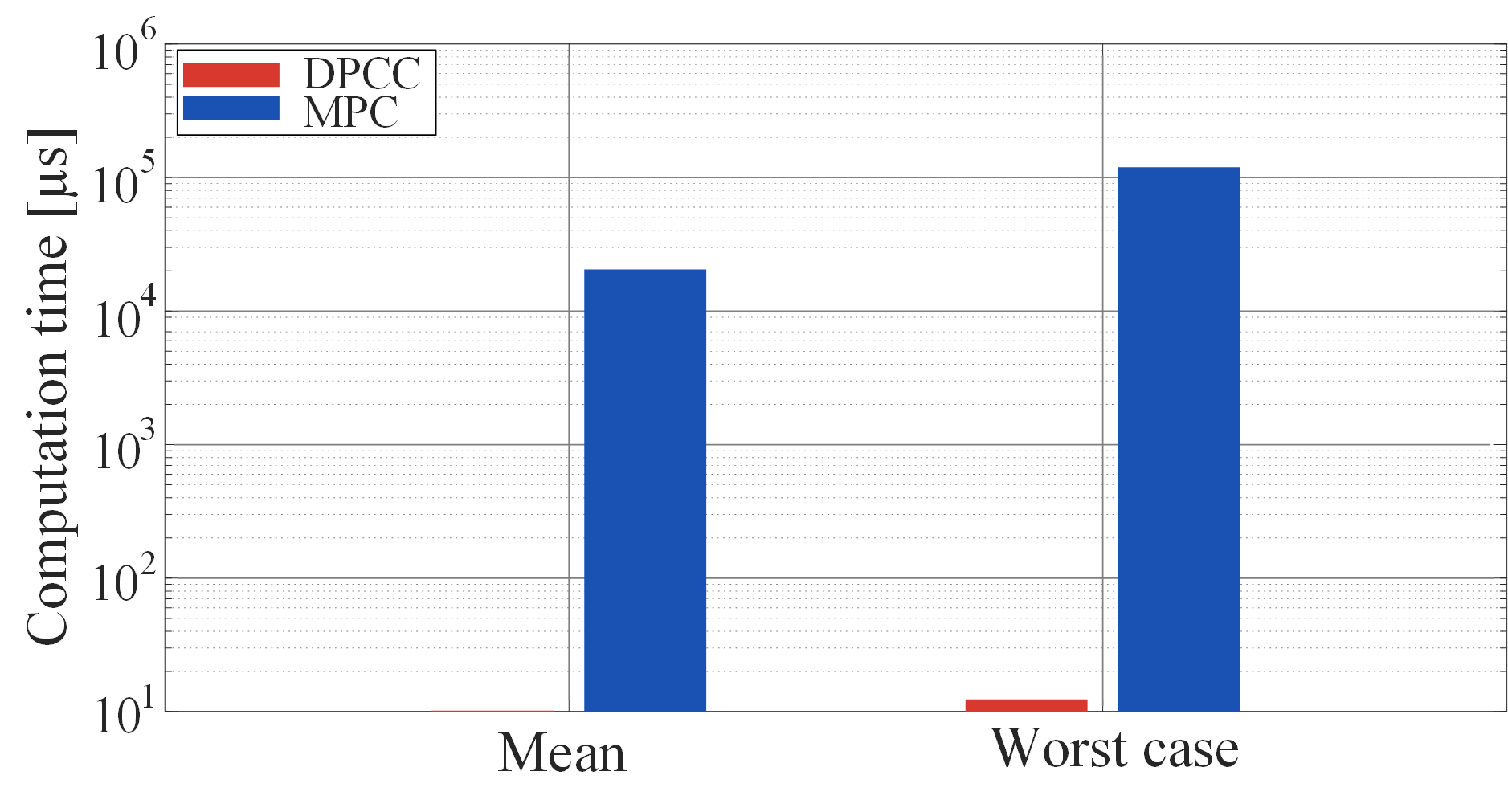}%
    }{%
        \figplaceholder{35mm}{\columnwidth}%
    }
    \caption{MATLAB computation time of \blueadd{DPCC inference and online MPC optimization}, including mean and worst-case values.}
    \label{fig:constrained_computation_time}
\end{figure}

Fig.~\ref{fig:constrained_computation_time} compares the
computation times of DPCC and online constrained MPC
over all 13,000 evaluation samples. The mean and
worst-case computation times are defined as the
average and maximum execution times, respectively.
DPCC requires approximately 10~$\mu$s on average
and 12~$\mu$s in the worst case, whereas the
corresponding values for MPC are approximately
$2\times10^{4}$ and $10^{5}$~$\mu$s, respectively.
The mean DPCC computation time is barely visible
on the logarithmic scale due to its small magnitude.
Together, Figs.~\ref{fig:constraint_violation_summary}--
\ref{fig:constrained_computation_time} demonstrate
that DPCC closely approximates the constrained MPC
policy while substantially reducing online
computational requirements.

\endgroup
\subsection{Sensorless Startup and Drive Operation}

Since the back EMF is weak at low speed and vanishes at standstill, the implemented drive uses a three-stage sensorless startup procedure consisting of rotor alignment, open-loop acceleration, and transition to back-EMF-based closed-loop operation. During alignment, a positive $d$-axis current is applied with $i_q^\star=0$. The motor is then accelerated in open loop using a limited $i_q^\star$, with the electrical angle generated as
\begin{equation}
\theta_e^{\mathrm{OL}}(k+1)
=
\theta_e^{\mathrm{OL}}(k)
+
T_s\omega_e^{\mathrm{OL}}(k).
\label{eq:open_loop_angle}
\end{equation}

Once sufficient speed is reached, the estimated $\alpha\beta$ back-EMF components are filtered and processed by a PLL to obtain $\hat{\theta}_e$ and $\hat{\omega}_e$ \cite{wu2023optimized_pll,zhang2024_discrete_time_pll}. The transition to sensorless closed-loop operation occurs at approximately 600~r/min, after which the estimated rotor position and speed are used for the coordinate transformations and DPCC.

\section{Experimental Validation}

\begin{table}[t]
\caption{Parameters of the Experimental PMSM Drive System}
\label{tab:motor_parameters}
\centering
\footnotesize
\setlength{\tabcolsep}{3pt}
\renewcommand{\arraystretch}{1.0}
\begin{tabular}{@{}p{0.39\columnwidth} c c c@{}}
\hline
\textbf{Parameter} & \textbf{Symbol} & \textbf{Value} & \textbf{Unit} \\
\hline
Pole pairs              & $p$             & 4             & -- \\
Rated power             & $P_N$           & 3.0           & kW \\
DC voltage              & $V_dc$           & 564           & V \\
Rated speed             & $n_N$           & 1500          & r/min \\
Rated torque            & $T_N$           & 19.1          & N$\cdot$m \\
Peak torque             & $T_{\mathrm{peak}}$ & 57.3       & N$\cdot$m \\
Rated current           & $I_N$           & 7.0           & A$_{\mathrm{rms}}$ \\
Rotor inertia           & $J$             & $1.07\times10^{-3}$ & kg$\cdot$m$^2$ \\
Line-to-line resistance & $R_{ll}$        & 3.08          & $\Omega$ \\
Line-to-line inductance & $L_{ll}$        & 37.1          & mH \\
BEMF at rated speed     & $E_N$           & 272           & V$_{\mathrm{rms}}$ \\
Torque constant         & $K_T$           & 2.74          & N$\cdot$m/A \\
Drive rated power       & $P_N$              & 4.0           & kW \\
PWM period              & $T_s$           & 100           & $\mu$s \\
\hline
\end{tabular}
\end{table}

The proposed DPCC is experimentally evaluated on an industrial PMSM
drive platform and directly compared with the conventional MPC
benchmark described in the previous sections. Particular attention is
paid to current transient performance, torque-disturbance response,
robustness against parameter mismatch, and real-time computational
requirements.

\begin{figure*}[!t]
    \centering
    \includegraphics[width=0.9\textwidth]{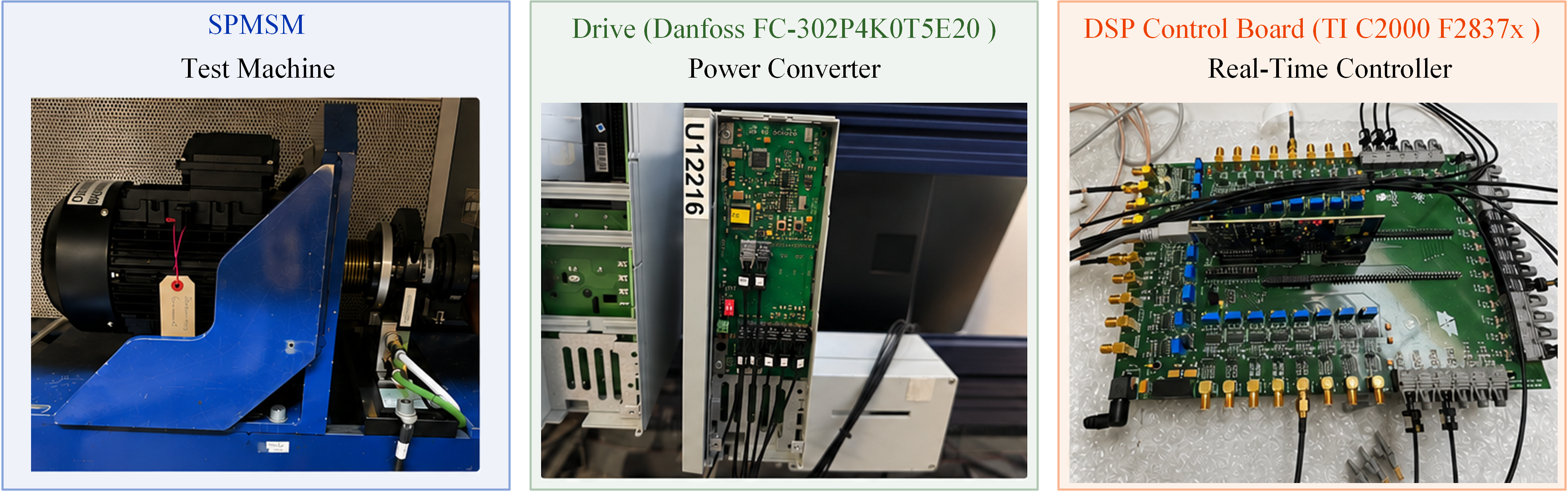}
    \caption{Industrial platform for the validation of the proposed DPCC method.}
    \label{fig:setup}
\end{figure*}

The experimental setup is shown in Fig.~\ref{fig:setup}.
The test bench consists of a surface-mounted PMSM, a commercial Danfoss
FC-302 frequency converter, and a TI C2000 F2837x DSP-based control
platform. The tested PMSM is an HPS90 1500 64 machine with the parameters listed in Table I. The DSP performs the real-time current-control calculations together
with PWM generation, signal acquisition, and communication with the
drive.

\subsection{Current Transient Performance}

\begin{figure*}[!t]
    \centering
    \begin{minipage}[t]{0.48\textwidth}
        \centering
        \includegraphics[width=\linewidth]{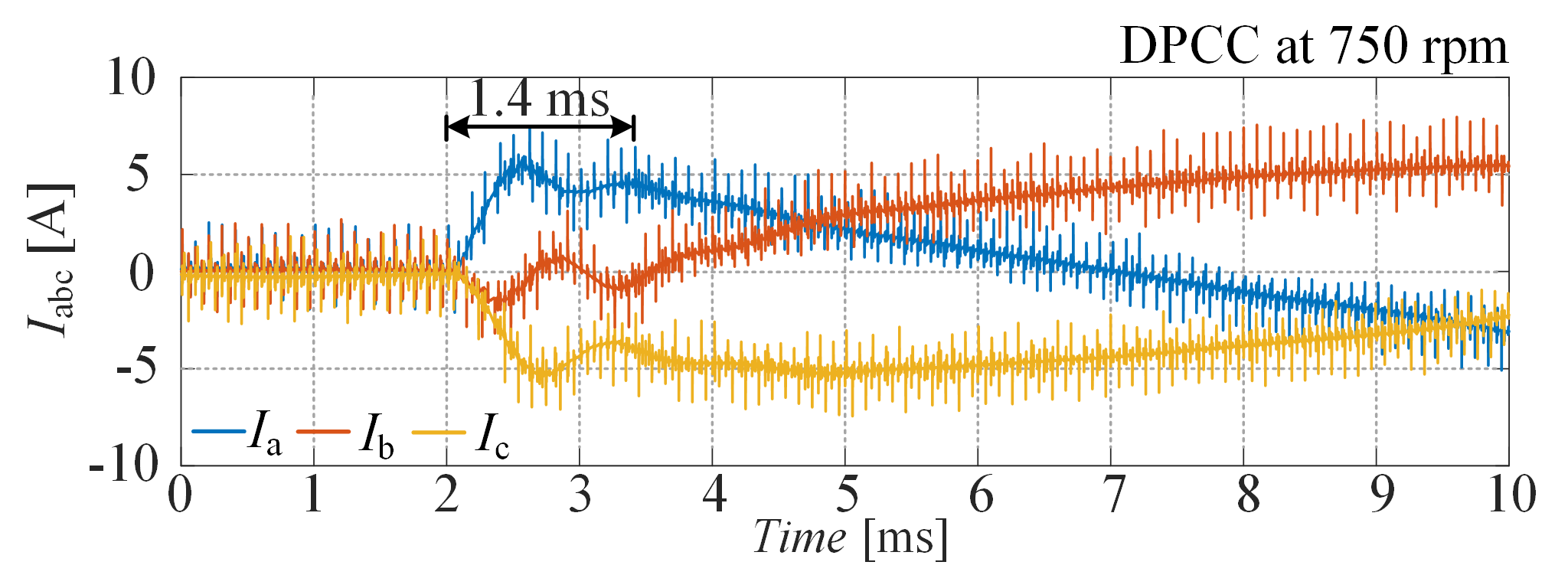}
        \centerline{(a)}
    \end{minipage}\hfill
    \begin{minipage}[t]{0.48\textwidth}
        \centering
        \includegraphics[width=\linewidth]{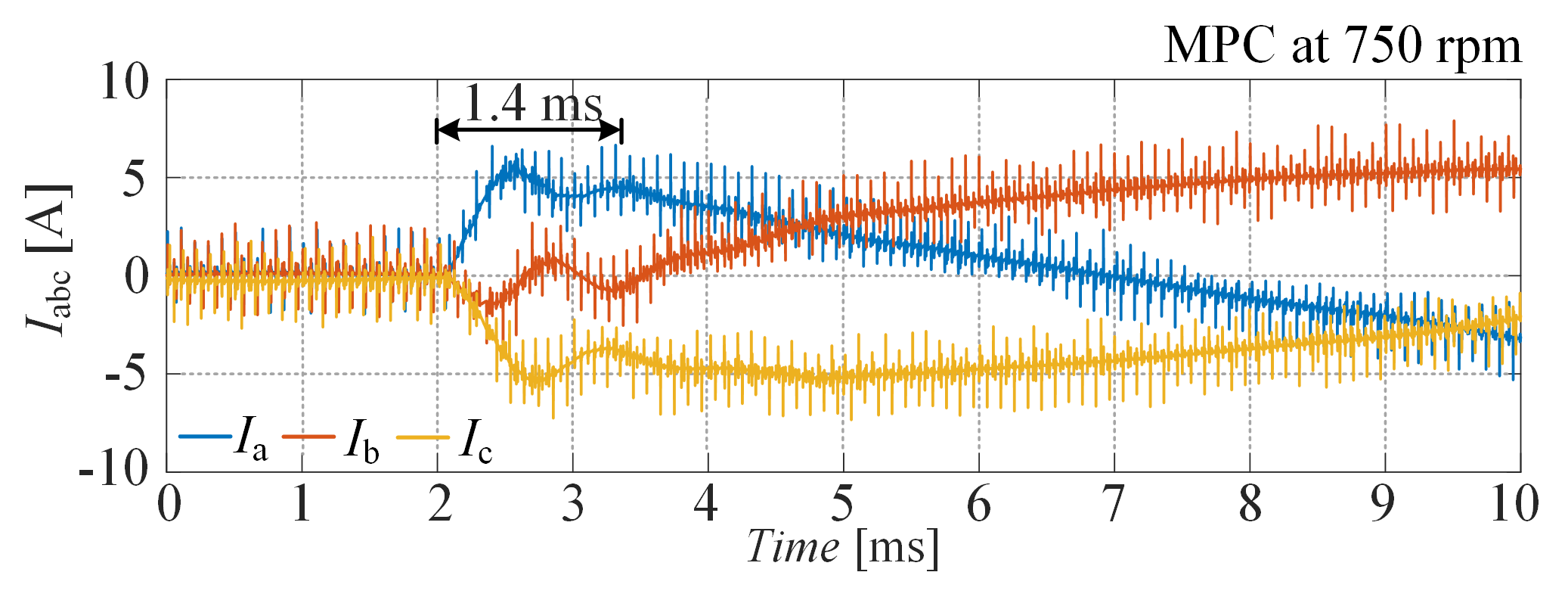}
        \centerline{(b)}
    \end{minipage}
    \vspace{1mm}
    \begin{minipage}[t]{0.48\textwidth}
        \centering
        \includegraphics[width=\linewidth]{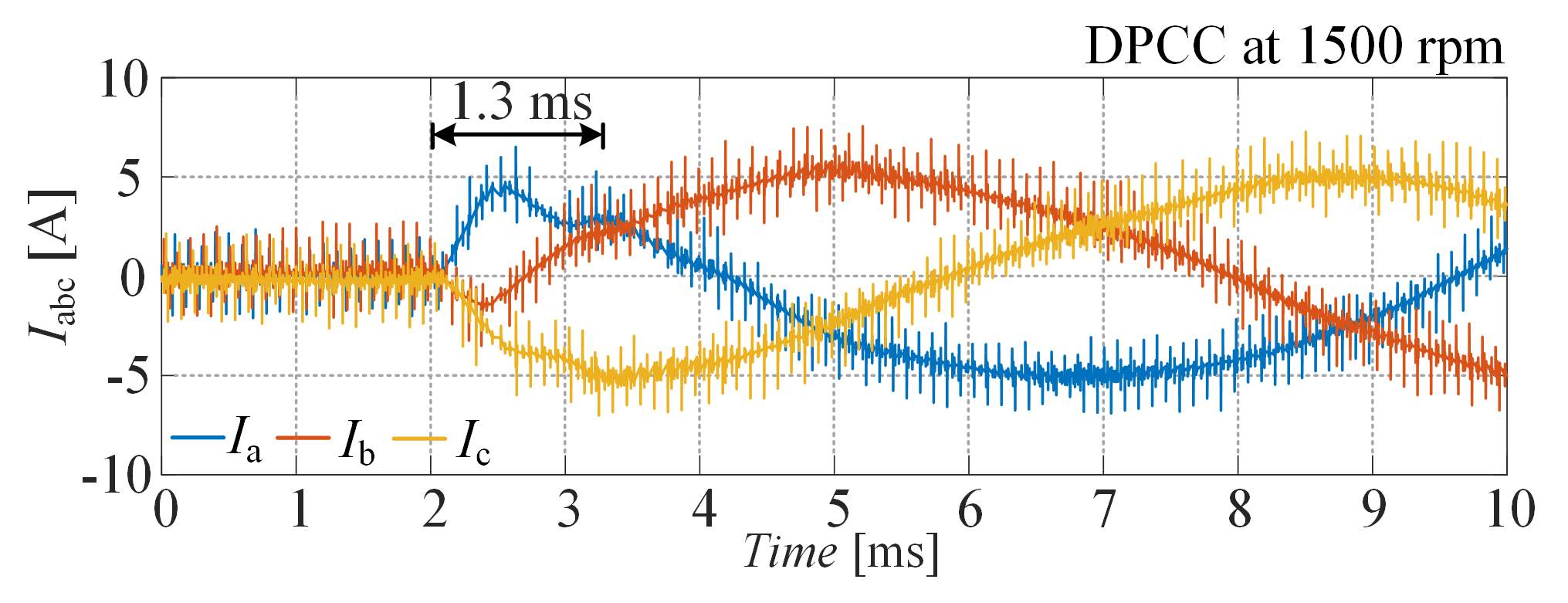}
        \centerline{(c)}
    \end{minipage}\hfill
    \begin{minipage}[t]{0.48\textwidth}
        \centering
        \includegraphics[width=\linewidth]{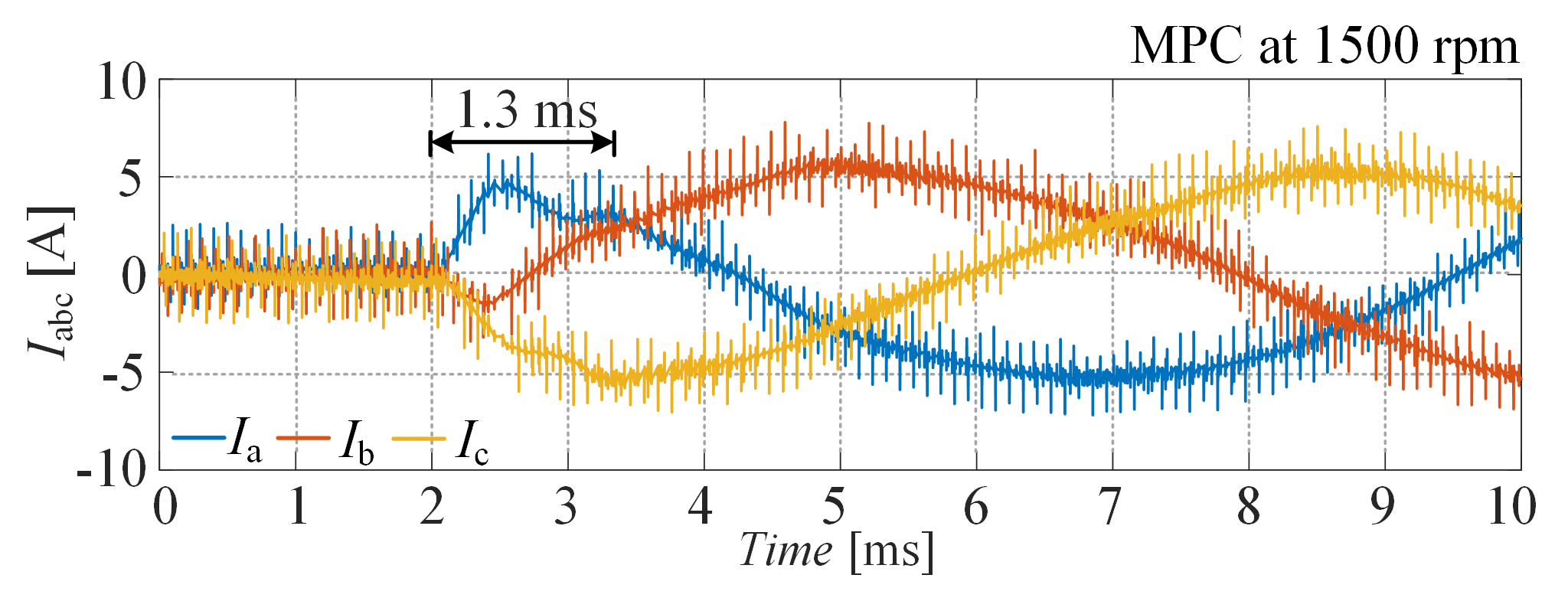}
        \centerline{(d)}
    \end{minipage}
    \caption{Measured experimental transient responses of the measured
    three-phase stator currents under a $0$--$5$~A current-reference
    transition with nominal machine parameters:
    (a) DPCC at 750~r/min;
    (b) MPC at 750~r/min;
    (c) DPCC at 1500~r/min; and
    (d) MPC at 1500~r/min.}
    \label{fig:current_05}
\end{figure*}

\begin{figure*}[!t]
    \centering
    \begin{minipage}[t]{0.48\textwidth}
        \centering
        \includegraphics[width=\linewidth]{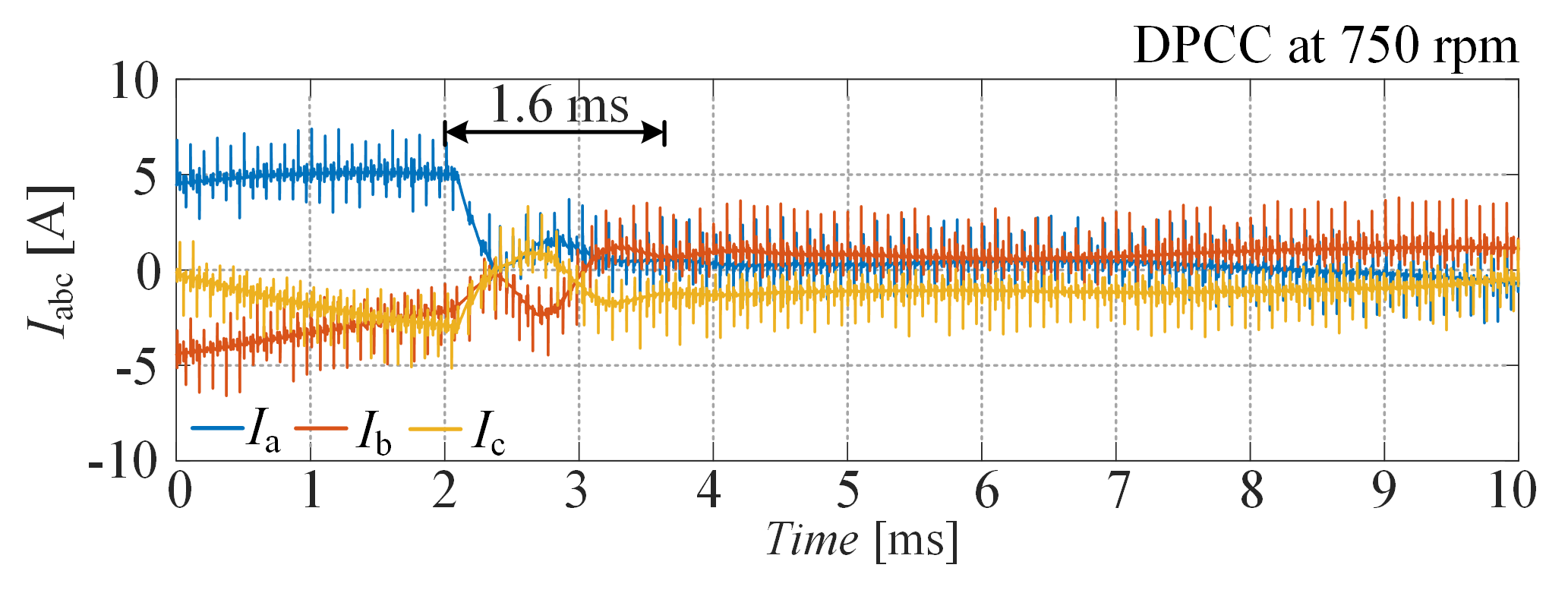}
        \centerline{(a)}
    \end{minipage}\hfill
    \begin{minipage}[t]{0.48\textwidth}
        \centering
        \includegraphics[width=\linewidth]{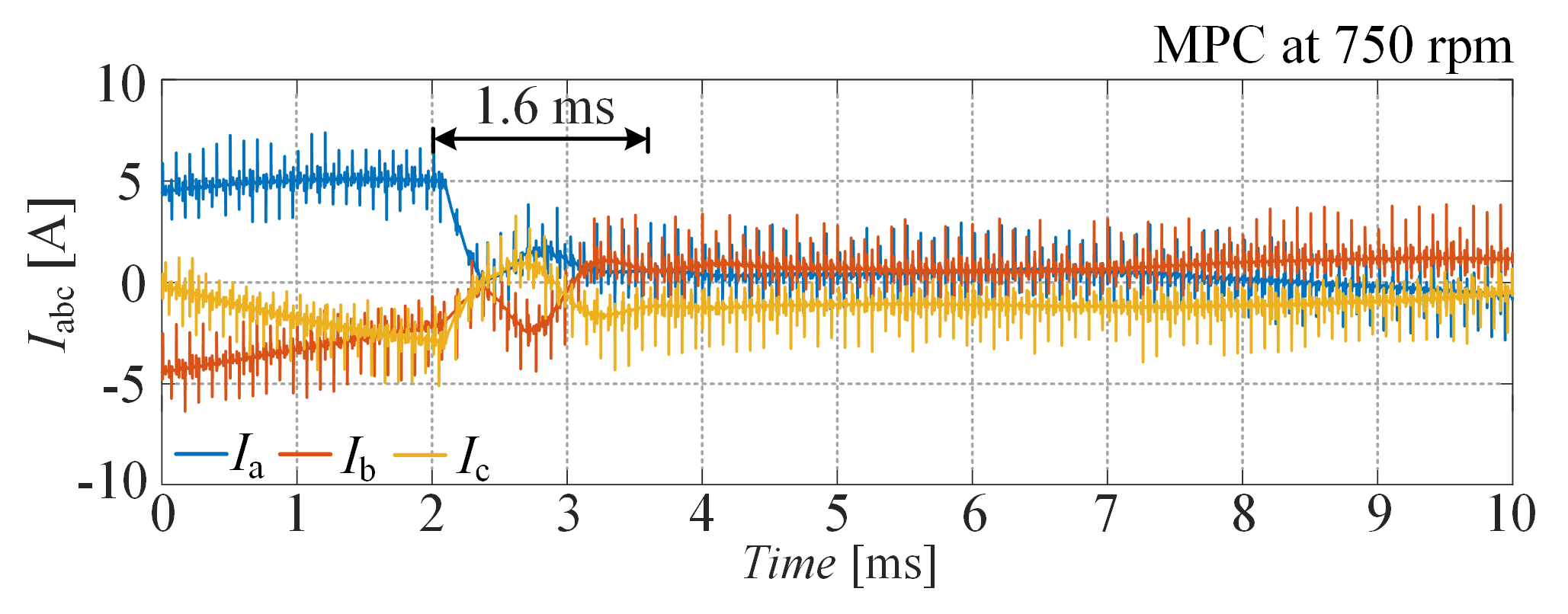}
        \centerline{(b)}
    \end{minipage}
    \vspace{1mm}
    \begin{minipage}[t]{0.48\textwidth}
        \centering
        \includegraphics[width=\linewidth]{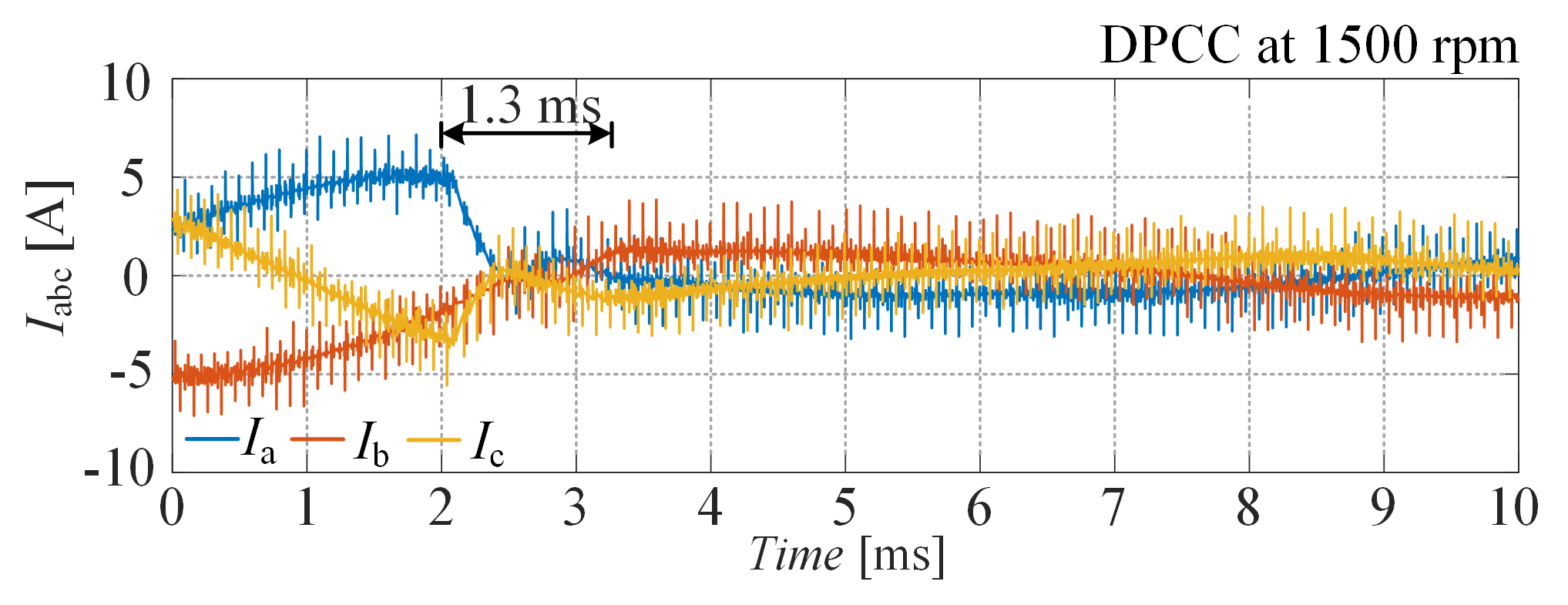}
        \centerline{(c)}
    \end{minipage}\hfill
    \begin{minipage}[t]{0.48\textwidth}
        \centering
        \includegraphics[width=\linewidth]{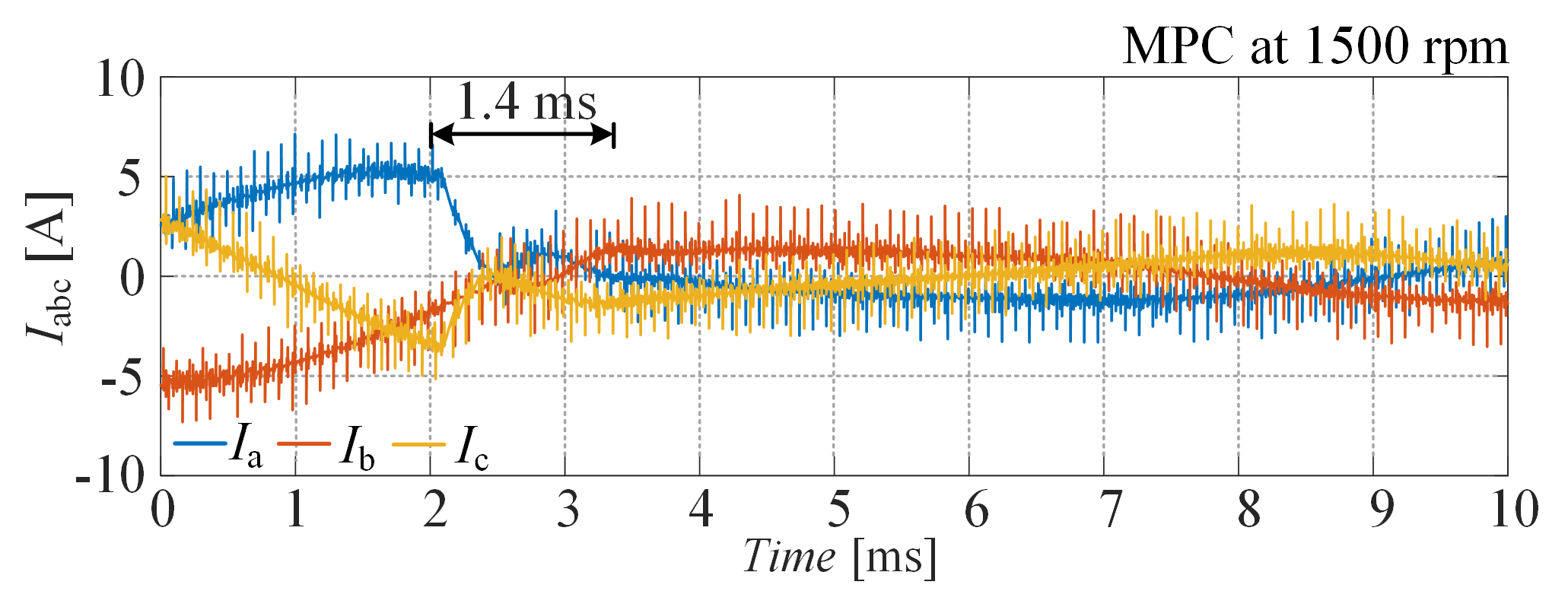}
        \centerline{(d)}
    \end{minipage}
    \caption{Measured experimental transient responses of the measured
    three-phase stator currents under a $5$--$1$~A current-reference
    transition with nominal machine parameters:
    (a) DPCC at 750~r/min;
    (b) MPC at 750~r/min;
    (c) DPCC at 1500~r/min; and
    (d) MPC at 1500~r/min.}
    \label{fig:current_51}
\end{figure*}

The current transient performance is first evaluated using the nominal
machine parameters. Two $d$-axis current reference changes, from
$0$ to $5$~A and from $5$ to $1$~A, are tested at 750~r/min and
1500~r/min. The corresponding experimental results for DPCC and MPC
are presented in Figs.~\ref{fig:current_05} and \ref{fig:current_51}. For each reference transition, the results at 750~r/min and
1500~r/min are grouped in one figure to facilitate direct comparison
between the two operating speeds and the two controllers.

The measured three phase stator currents are presented directly,
without reconstruction of the $dq$ currents using the estimated
electrical angle. After each current reference change, a short
transient appears in the phase currents before the periodic current
waveforms corresponding to the new operating point are established.
High frequency current ripple is observed during the transient due to
the switching operation. 

For the $0$ to $5$~A current reference change, the transient interval
is approximately 1.4~ms for both DPCC and MPC at 750~r/min, as shown
in Fig.~\ref{fig:current_05}(a) and (b). At 1500~r/min, the corresponding
value is approximately 1.3~ms for both controllers, as shown in Fig.~\ref{fig:current_05}(c) and (d). The results at both operating speeds show that DPCC provides a transient response comparable to that of MPC.

Similar results are obtained for the $5$ to $1$~A current reference
change. At 750~r/min, the transient interval is approximately 1.6~ms
for both controllers, as shown in Fig.~\ref{fig:current_51}(a) and (b). At
1500~r/min, the transient intervals are approximately 1.3~ms for DPCC
and 1.4~ms for MPC, as shown in Fig.~\ref{fig:current_51}(c) and (d). The
result does not indicate
a significant difference in transient performance.

It should be noted that the phase current waveforms have a longer
fundamental period at 750~r/min because of the lower electrical
frequency. This difference in waveform period does not represent a
slower current control response. For all tested conditions, the
measured transient intervals remain within approximately
1.3--1.6~ms for both DPCC and MPC.

Overall, the experiments with nominal machine parameters show that
DPCC achieves current transient performance comparable to MPC at both
750~r/min and 1500~r/min, for both increasing and decreasing current
reference changes. These results indicate that replacing the online
MPC optimization with neural network inference does not cause a
measurable deterioration in the current transient response.

\subsection{Robustness Against Inductance Mismatch}

\begin{figure*}[!t]
    \centering
    \begin{minipage}[t]{0.48\textwidth}
        \centering
        \includegraphics[width=\linewidth]{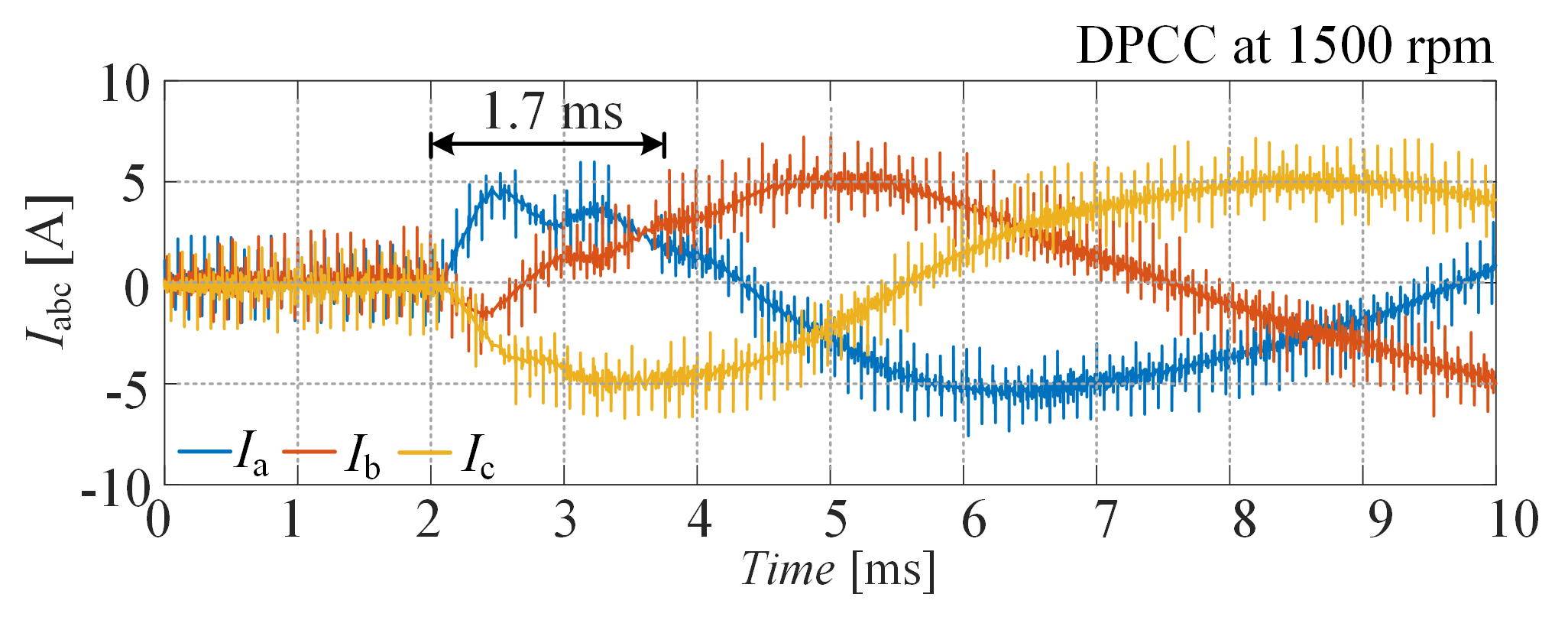}
        \centerline{(a)}
    \end{minipage}\hfill
    \begin{minipage}[t]{0.48\textwidth}
        \centering
        \includegraphics[width=\linewidth]{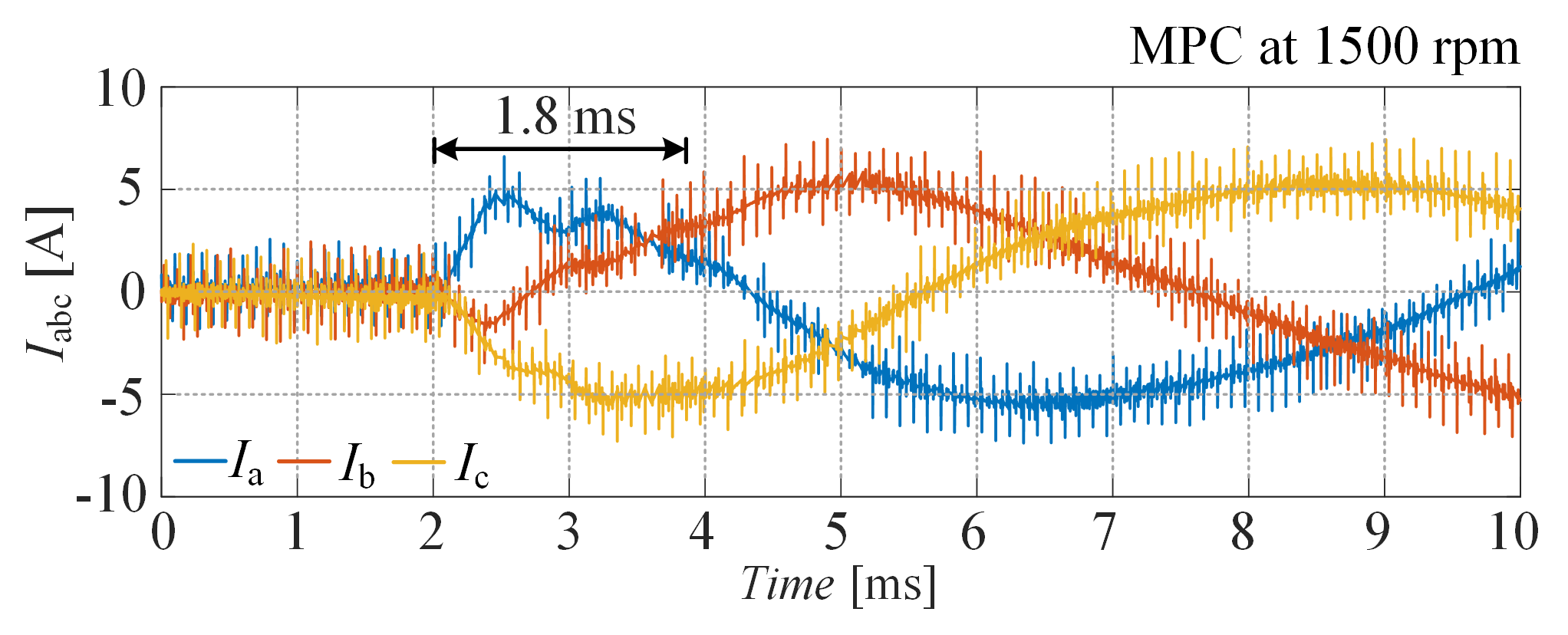}
        \centerline{(b)}
    \end{minipage}
    \vspace{1mm}
    \begin{minipage}[t]{0.48\textwidth}
        \centering
        \includegraphics[width=\linewidth]{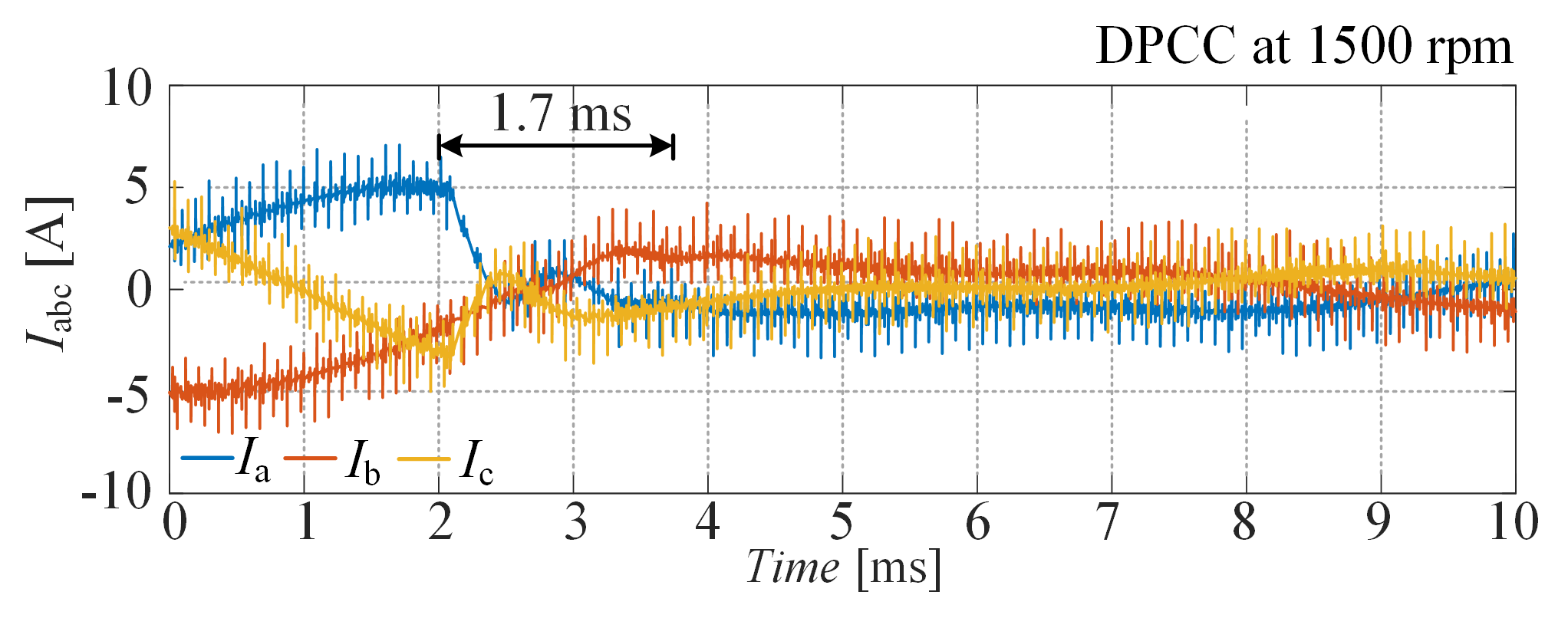}
        \centerline{(c)}
    \end{minipage}\hfill
    \begin{minipage}[t]{0.48\textwidth}
        \centering
        \includegraphics[width=\linewidth]{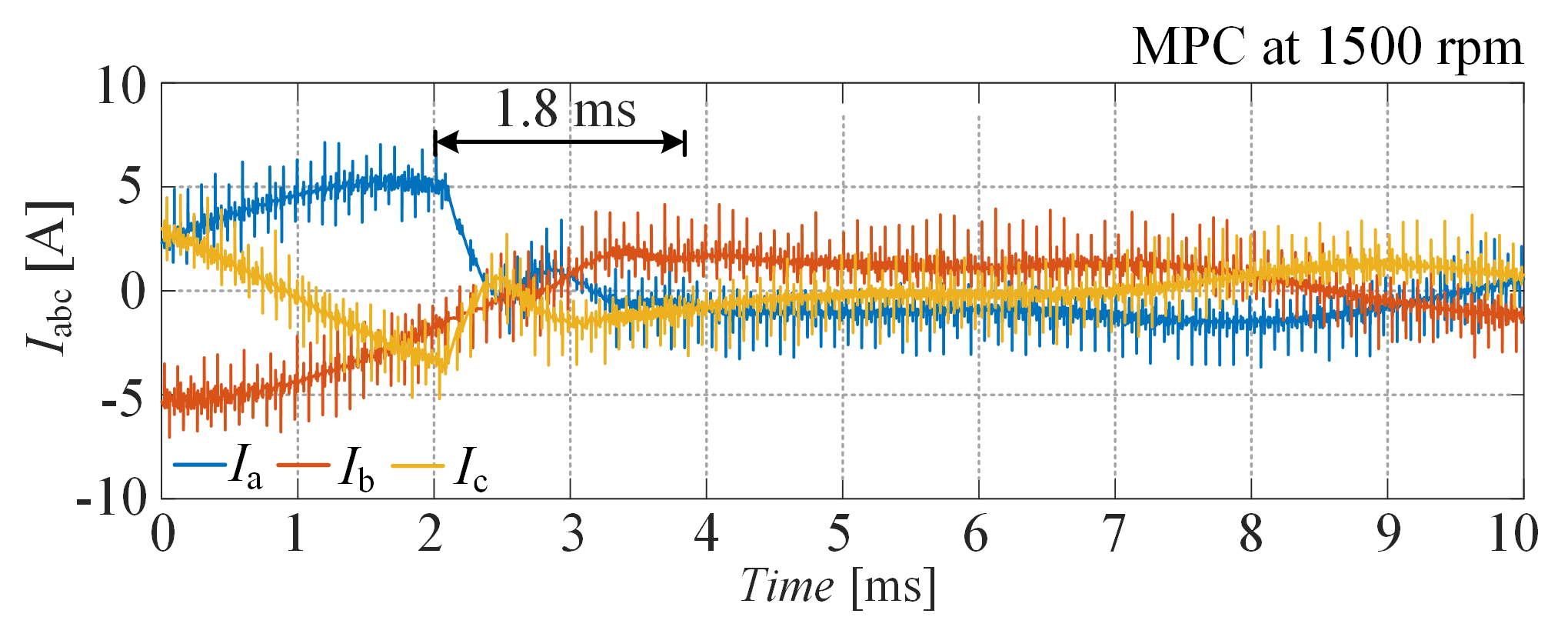}
        \centerline{(d)}
    \end{minipage}
    \caption{Measured experimental three-phase current responses with a
    20\% inductance mismatch at 1500~r/min:
    (a) DPCC, $0$--$5$~A;
    (b) MPC, $0$--$5$~A;
    (c) DPCC, $5$--$1$~A; and
    (d) MPC, $5$--$1$~A.}
    \label{fig:inductance_mismatch}
\end{figure*}

To evaluate robustness against inductance mismatch, the inductance used in both the DPCC training model and the MPC model is set to $120\%$ of its nominal value, while the actual motor inductance remains at the nominal value. The tests are conducted at the rated speed of 1500~r/min for both the $0$--$5$~A and $5$--$1$~A current-reference transitions, and the corresponding DPCC and MPC responses are shown in Fig.~\ref{fig:inductance_mismatch}. Compared with the nominal-parameter case, the inductance mismatch results in larger transient deviations and slower convergence following the current-reference transitions. Nevertheless, both DPCC and MPC maintain stable current regulation and converge to the corresponding steady-state current trajectories. No sustained oscillations, excessive current excursions, or loss of current regulation are observed.

For DPCC, the dominant transient intervals under the 20$\%$ inductance mismatch are approximately 1.7~ms for the $0$--$5$~A transition and 1.7~ms for the $5$--$1$~A transition, compared with approximately 1.3~ms under nominal parameters at the same operating speed. The increased transient duration reflects the expected dependence of the model-based learned policy on parameter accuracy; however, stable closed-loop operation is preserved despite the imposed mismatch. MPC shows a comparable degradation in transient response under the same mismatch condition, confirming that the inductance deviation affects both model-dependent control approaches.

\subsection{Load-Torque Response}

\begin{figure}[!t]
    \centering
    \includegraphics[width=1\columnwidth]{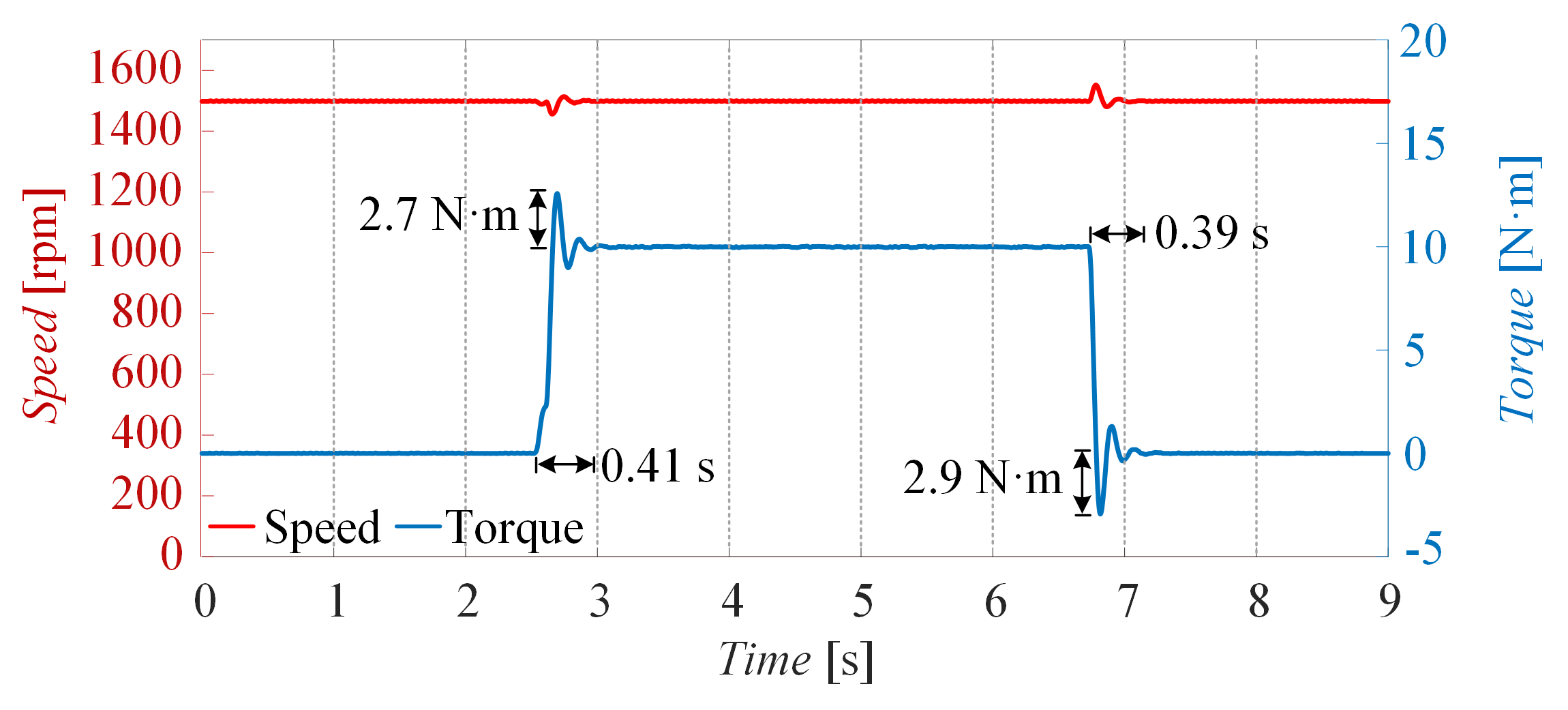}
    \centerline{(a)}
    \vspace{2mm}
    \includegraphics[width=1\columnwidth]{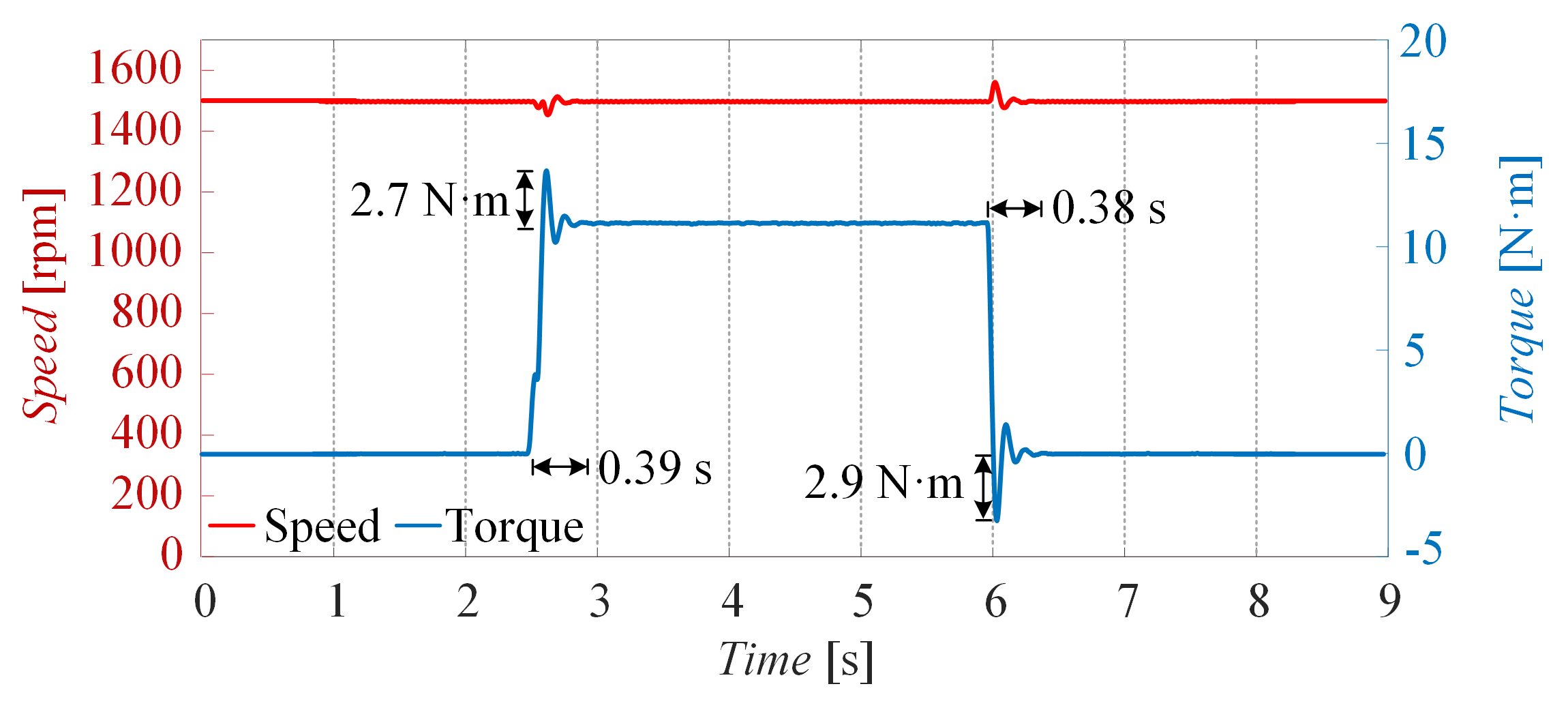}
    \centerline{(b)}
    \caption{Experimental speed and torque responses to a
    10~N$\cdot$m load-torque step at approximately 1500~r/min:
    (a) proposed DPCC and (b) conventional MPC.}
    \label{fig:torque_step}
\end{figure}

The torque response is evaluated by applying
a 10~N$\cdot$m load-torque step while the motor operates at 1500~r/min. The measured speed and torque responses of
DPCC and MPC are shown in Fig.~\ref{fig:torque_step}.

Before the disturbance is applied, the motor operates at the rated speed under the no-load condition. When the load torque is
increased to approximately 10~N$\cdot$m, both controllers rapidly
develop the required electromagnetic torque. DPCC reaches the new
loaded operating condition within approximately 0.41~s, while MPC
requires approximately 0.40~s. A short torque excursion of
approximately 2.7~N$\cdot$m relative to the new steady-state level is
observed for both controllers during load application. At the same
time, only a small and short-duration speed deviation occurs, after
which the speed returns close to its reference.

When the load is subsequently removed, both controllers again exhibit
closely matched responses. The transient torque deviation is
approximately 2.9~N$\cdot$m, and the original no-load operating
condition is recovered within approximately 0.39~s for both DPCC and
MPC. The close correspondence during both load application and load
removal indicates that the neural-network implementation preserves the characteristics of the MPC controller.

\subsection{Real-Time Computational Performance}

\begin{figure}[!t]
    \centering
    \includegraphics[width=\columnwidth]{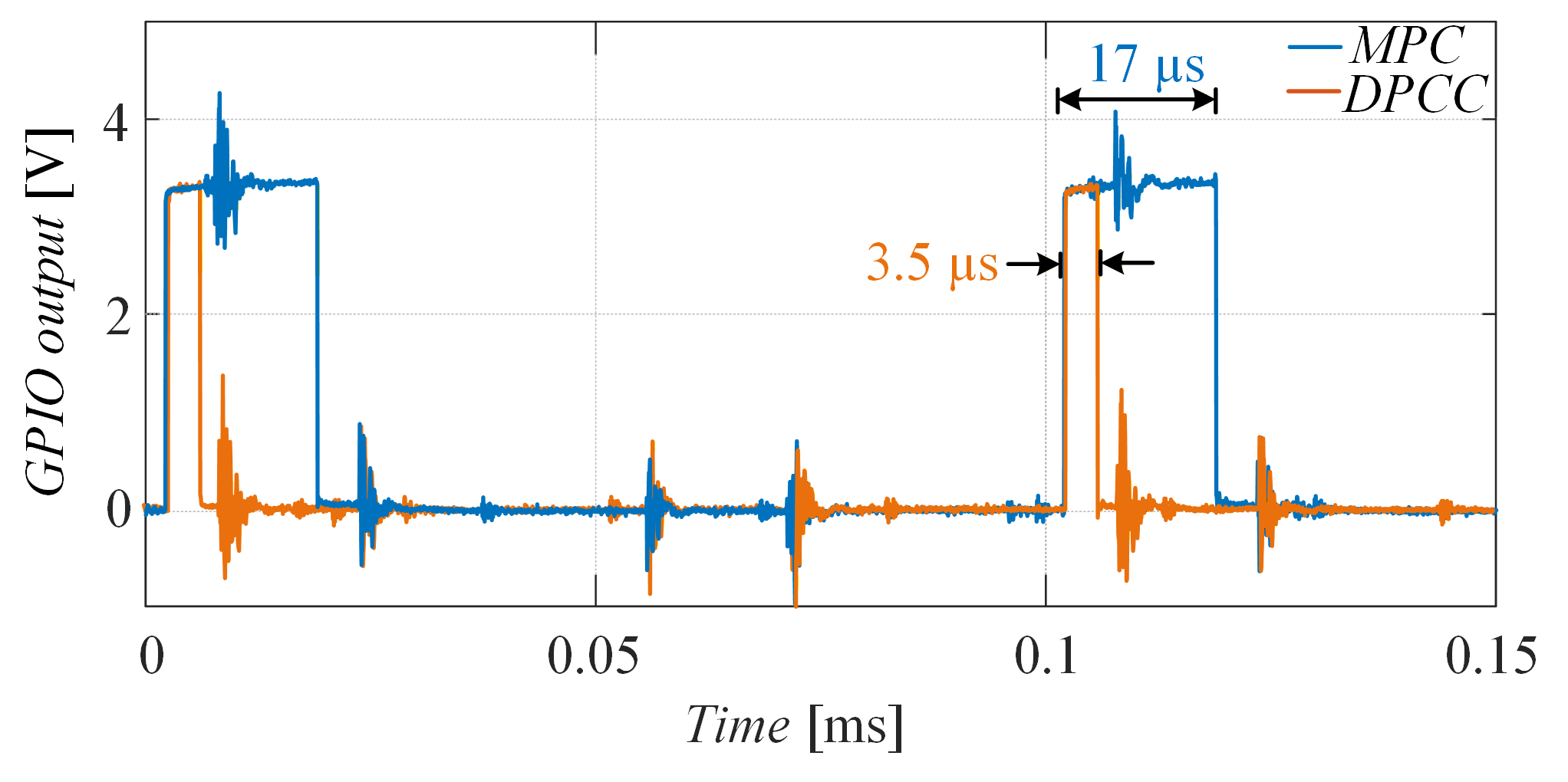}
    \caption{Measured GPIO timing signals for the DPCC and optimized
    symbolic MPC implementations on the TI C2000 F2837x DSP.}
    \label{fig:execution_time}
\end{figure}

Real-time computational efficiency is one of the primary motivations
for replacing online MPC optimization with DPCC. The MPC benchmark employs the optimized symbolic implementation
described in Section~\ref{sec:symbolic_mpc}, where the matrix
factorization and algebraic expansion are completed offline and only
the resulting scalar expressions are evaluated online. 

The execution time is measured using a dedicated GPIO signal that is
activated during the execution of the current-control routine. Therefore, execution-time evaluation is necessary to verify whether the controller can satisfy the sampling-time constraints of embedded motor-drive platforms \cite{cortes2012delay_compensation,gao2020delay_pmsm}. The pulse
width therefore provides a direct hardware measurement of the
controller computation time. As shown in Fig.~\ref{fig:execution_time},
DPCC reduces the execution time by approximately 79\% relative to the
optimized analytical MPC implementation. This reduction is achieved even
though the MPC benchmark does not use an online general-purpose solver
or matrix inversion.

In summary, the experimental results show that DPCC achieves
current and torque responses comparable to MPC while requiring
substantially less real-time computation. Its stable operation under
inductance mismatch and integration into the complete sensorless
industrial drive further support its feasibility for embedded PMSM
control.

\section{Conclusion}

This paper proposed a differentiable predictive current control (DPCC) method for PMSM drives that transfers the predictive optimization of MPC to offline differentiable training. By embedding the PMSM model and predictive current-tracking objective into the training process, the neural policy is optimized directly from the control objective without MPC-generated labels, reducing online control to a fixed-cost neural-network inference.

The proposed DPCC was implemented using a lightweight $5$--$6$--$6$--$2$ neural network on an industrial PMSM drive platform with a practical sensorless architecture. Experimental results under current transients, load-torque disturbances, and inductance mismatch show performance comparable to the offline-expanded analytical MPC benchmark, while reducing the execution time to 3.5~$\mu$s. These results demonstrate the potential of DPCC to retain MPC-like control performance with substantially lower online computation for industrial PMSM drives.

\section*{Acknowledgment}
This work was supported by the Innovation Fund Denmark (IFD) through the project of Artificial Intelligence for Next-Generation Power Electronics (AI-Power).

\bibliographystyle{IEEEtran}
\bibliography{IEEEabrv,references}

\end{document}